\documentclass[conference]{IEEEtran}
\IEEEoverridecommandlockouts
\usepackage{cite}
\usepackage{amsmath,amssymb,amsfonts}
\usepackage{algorithmic}
\usepackage{graphicx}
\usepackage{textcomp}
\usepackage{xcolor}
\usepackage{float}
\usepackage{array}
\usepackage{multirow}
\usepackage[ruled,vlined]{algorithm2e} 
\usepackage{url}

\usepackage{fancyhdr}

\def\BibTeX{{\rm B\kern-.05em{\sc i\kern-.025em b}\kern-.08em
    T\kern-.1667em\lower.7ex\hbox{E}\kern-.125emX}}
\begin{document}

\title{Explainable Transformer-CNN Fusion for Noise-Robust Speech Emotion Recognition\\
\thanks{AmygdalaAI-India Lab, is an international volunteer-run research group that advocates for \textit{AI for a better tomorrow} https://amygdalaaiindia.github.io/.}
}

\author{\IEEEauthorblockN{Sudip Chakrabarty, Pappu Bishwas, Rajdeep Chatterjee}

\IEEEauthorblockA{School of Computer Engineering, KIIT Deemed to be University, Bhubaneswar 751024, India\\
Amygdala-AI India Lab, Bhubaneswar, India\\
Email: \{sudipchakrabarty6, pappuovi8, cse.rajdeep\}@gmail.com}}


\maketitle

\begin{abstract}
Speech Emotion Recognition (SER) systems often degrade in performance when exposed to the unpredictable acoustic interference found in real-world environments. Additionally, the opacity of deep learning models hinders their adoption in trust-sensitive applications. To bridge this gap, we propose a Hybrid Transformer-CNN framework that unifies the contextual modeling of Wav2Vec 2.0 with the spectral stability of 1D-Convolutional Neural Networks. Our dual-stream architecture processes raw waveforms to capture long-range temporal dependencies while simultaneously extracting noise-resistant spectral features (MFCC, ZCR, RMSE) via a custom Attentive Temporal Pooling mechanism. We conducted extensive validation across four diverse benchmark datasets: RAVDESS, TESS, SAVEE, and CREMA-D. To rigorously test robustness, we subjected the model to non-stationary acoustic interference using real-world noise profiles from the SAS-KIIT dataset. The proposed framework demonstrates superior generalization and state-of-the-art accuracy across all datasets, significantly outperforming single-branch baselines under realistic environmental interference. Furthermore, we address the ``black-box" problem by integrating SHAP and Score-CAM into the evaluation pipeline. These tools provide granular visual explanations, revealing how the model strategically shifts attention between temporal and spectral cues to maintain reliability in the presence of complex environmental noise.
\end{abstract}


\vspace{-1mm}
\section*{\textbf{List of Abbreviations}}
\vspace{-1.5mm}
\begin{table}[h]
\centering
\begin{tabular}{|l|l|}
\hline
\textbf{Abbreviation} & \textbf{Definition} \\
\hline
SER & Speech Emotion Recognition \\
CNN & Convolutional Neural Network \\
MFCC & Mel-Frequency Cepstral Coefficients \\
ZCR & Zero-Crossing Rate \\
RMSE & Root Mean Square Energy \\
SNR & Signal-to-Noise Ratio \\
XAI & Explainable Artificial Intelligence \\
SHAP & SHapley Additive exPlanations \\
Score-CAM & Score-Weighted Class Activation Mapping \\
RAVDESS & Ryerson Audio-Visual Database of Emotional Speech \\
& and Song \\
TESS & Toronto Emotional Speech Set \\
CREMA-D & Crowd-sourced Emotional Multimodal Actors Dataset \\
SAVEE & Surrey Audio-Visual Expressed Emotion\\
SAS-KIIT & South Asian Sound Dataset - KIIT\\
IEMOCAP & Interactive Emotional Dyadic Motion Capture\\
STFT & Short-Time Fourier Transform \\
DCT & Discrete Cosine Transform \\
PSD & Power Spectral Density \\
GELU & Gaussian Error Linear Unit \\
HMM & Hidden Markov Model\\

\hline
\end{tabular}
\label{tab:abbreviations}
\end{table}
\vspace{-2mm}

\section{\textbf{Introduction}}
\label{sec:introduction}
Speech Emotion Recognition (SER) has emerged as a pivotal component in the advancement of Human-Computer Interaction (HCI), bridging the gap between digital systems and human affective states \cite{al2023speech}. The ability to accurately decode emotional intent from audio signals enables transformative applications, ranging from AI-driven mental health monitoring and automated customer service analysis to adaptive e-learning environments \cite{jordan2025speech,9352018}. As voice-enabled assistants become ubiquitous, the demand for SER systems that can operate reliably in diverse acoustic settings has never been higher. However, despite the impressive performance of deep learning models in controlled environments, deploying these systems in the real world remains a significant engineering challenge.

The primary obstacle hindering widespread adoption is environmental robustness. State-of-the-art models often achieve near-perfect accuracy on clean, studio-recorded datasets but suffer catastrophic performance degradation when exposed to non-stationary noise, such as traffic sounds, background chatter, or machinery hum \cite{10887569}. This ``domain shift'' between training and deployment creates a fragility that renders many high-performing models unusable in the wild. Furthermore, the increasing complexity of Deep Neural Networks (DNNs) has introduced a secondary crisis: the \textbf{``Black Box''} problem \cite{arrieta2020explainable, adadi2018peeking, rudin2019stop, 11205342}. In sensitive domains like healthcare or security, a model’s prediction alone is insufficient; stakeholders require transparent reasoning behind the decision. The opacity of modern architectures, where millions of parameters interact in non-linear ways, erodes user trust and complicates the debugging of failure modes in noisy conditions \cite{norval2025explainable}.

Existing literature has attempted to mitigate these issues through various architectural paradigms, yet limitations persist. Traditional approaches relying on handcrafted features (e.g., MFCCs, chroma) offer interpretability and spectral stability but fail to capture the high-level semantic context required for subtle emotion differentiation \cite{AOUANI2020251}. Conversely, deep Convolutional Neural Networks (CNNs) excel at extracting local patterns from spectrograms but often struggle to model long-range temporal dependencies. More recently, Transformer-based architectures, such as Wav2Vec 2.0, have set new benchmarks by leveraging self-attention mechanisms to model global context \cite{baevski2020wav2vec}. However, purely Transformer-based approaches can be computationally intensive and may overlook fine-grained spectral cues that are crucial when the linguistic content is masked by noise. There remains a critical need for an architecture that can synergize the temporal power of Transformers with the spectral robustness of CNNs while remaining interpretable to human observers.

To address these limitations, this paper proposes a unified Dual-Stream Fusion Network. Unlike prior works that rely on a single modality, our approach strategically integrates the long-range dependency modeling of Wav2Vec 2.0 with noise-invariant spectral features. Furthermore, we explicitly tackle the ``black box" transparency issue by integrating explainable AI modules directly into the recognition pipeline. The main contributions of this work are summarized as follows:

\begin{itemize}
    \item \textbf{Synergistic Dual-Stream Architecture:} We propose a novel fusion framework that strategically bridges the gap between deep semantic modeling and low-level signal stability. By integrating a fine-tuned \textbf{Wav2Vec 2.0} backbone with a lightweight \textbf{1D-CNN} via our custom Attentive Temporal Pooling, the model effectively captures long-range linguistic context while maintaining robustness against spectral corruption.

    \item \textbf{Resilience to Real-World Acoustic Degradation:} We rigorously validate the model's stability against a spectrum of environmental interferences, ranging from stationary profiles (White, Pink noise) to highly non-stationary, chaotic soundscapes (using the \textbf{SAS-KIIT} dataset). The results demonstrate that our fusion approach significantly outperforms single-branch baselines in high-intensity noise scenarios.

    \item \textbf{Cross-Corpus Generalization:} To ensure distinct demographic and acoustic universality, the framework is evaluated across four diverse benchmark corpora (\textbf{RAVDESS, TESS, SAVEE, CREMA-D}). This validates the model's ability to generalize across varying accents, speaker genders, and recording environments without overfitting.

    \item \textbf{Interpretable Decision-Making (XAI):} Addressing the "black-box" limitations of deep SER systems, we integrate a comprehensive Explainable AI suite using \textbf{SHAP} and \textbf{Score-CAM}. These tools provide granular visual evidence that the model correctly prioritizes linguistically significant acoustic segments (e.g., prosodic onsets) rather than relying on background noise artifacts.
\end{itemize}

The remainder of this paper is organized as follows. Section \ref{sec:related_work} reviews the existing literature on speech emotion recognition, focusing on noise-robustness techniques. Section \ref{sec:methodology} delineates the proposed dual-stream framework, detailing the Wav2Vec 2.0 backbone, the CNN branch, and the attentive fusion strategy. Section \ref{sec:results_analysis} presents the quantitative performance evaluation, including comprehensive robustness benchmarks and component ablation studies. Section \ref{sec:xai} is dedicated to model transparency, offering a detailed Explainable AI analysis to validate decision-making. Section \ref{sec:discussion} interprets these findings in the context of acoustic volatility and architectural synergy. Finally, Section \ref{sec:conclusion} summarizes the key contributions and outlines potential directions for future research.

\section{\textbf{Related Works}}
\label{sec:related_work}
Existing literature in Speech Emotion Recognition can be broadly categorized into three streams: spectral feature-based approaches, end-to-end deep learning models, and hybrid fusion architectures \cite{10711189}. While significant progress has been made in improving classification accuracy on standard datasets like RAVDESS \cite{revdess} and IEMOCAP\cite{busso2008iemocap}, two critical challenges remain prevalent: acoustic robustness in non-stationary environments and model interpretability. Speech emotion recognition has evolved significantly over the last two decades \cite{10.1145/3129340, kaur2023trends, 8757528}. Tzeng et al. proposed a unified framework integrating speech enhancement with self-supervised representations, achieving significant F1 gains at 0 dB SNR via a cascade unfrozen training strategy \cite{10887569}. However, their reliance on explicit enhancement increases computational overhead and risks suppressing subtle paralinguistic cues (e.g., breathiness) crucial for distinguishing high-arousal emotions in real-world scenarios. While prior research \cite{11173395} in audio analysis has prioritized reducing model complexity and accelerating inference speeds, these efficiency-driven approaches often entrench the `black box' paradigm, sacrificing internal transparency for computational performance. Nfissi et al. utilized SHAP to quantify feature importance, enhancing interpretability for standard SER benchmarks \cite{nfissi2024unveiling}. However, their analysis relies on clean data, leaving a critical gap in understanding how non-stationary noise distorts model attention, a dynamic we explicitly visualize in this study. A recent study \cite{sharma2025speech} has employed Local Interpretable Model-agnostic Explanations (LIME) to enhance transparency in SER, using local linear approximations to identify critical spectral features (e.g., MFCCs, CQT) for clinical applications. However, LIME \cite{ribeiro2016should} is frequently criticized for its computational instability, where the random perturbation process yields inconsistent explanations for identical inputs, a limitation our game-theoretic SHAP approach explicitly resolves. Another study \cite{app14041553} applied Grad-CAM \cite{selvaraju2017grad} to validate that models focus on relevant acoustic segments. However, its reliance on backpropagated gradients makes it susceptible to saturation in deep architectures, a limitation our gradient-free Score-CAM avoids. Furthermore, while Transformers have demonstrated superior context modeling, their susceptibility to background noise remains an active area of investigation. As demonstrated in Table \ref{tab:related_works}, the landscape of noise-robust SER has been dominated by classical classifiers (SVM, HMM) and standard deep learning architectures (CNN, LSTM). However, the utilization of Transformer-based models remains significantly underexplored in adverse acoustic conditions. Chen et al. \cite{chen2023noise} integrated a HuBERT Transformer with a CMGAN-based speech enhancement module to clean noisy audio [10]. However, this heavy generative denoising was found to suppress signal intensity, leading to reduced performance in detecting Arousal and Dominance, a limitation our feature fusion approach avoids by preserving raw prosodic energy. A detailed comparison of recent state-of-the-art techniques, along with their architectural focus and identified gaps, is presented in Table \ref{tab:related_works}.

\begin{table*}[ht!]
\centering
\caption{Summary of Related Works in Speech Emotion Recognition under Noisy Conditions}
\label{tab:related_works}
\renewcommand{\arraystretch}{1.3}
\begin{tabular}{|p{0.4cm}|p{0.6cm}|p{2.7cm}|p{1.5cm}|p{2cm}|p{3.8cm}|p{3.8cm}|}
\hline
\textbf{Ref.} & \textbf{Year} & \textbf{Method / Model} & \textbf{Dataset(s)} & \textbf{Noise / Condition} & \textbf{Key Findings / Accuracy} & \textbf{Notes / Limitation} \\

\hline
\cite{win2020} & 2020 & MLP / SVM / Decision Tree / Deep Learning & IEMOCAP & Real-world sport event noise (0–15 dB SNR) & MLP: 90.68\%, SVM: 56.52\%, Decision Tree: 89.44\%, Deep Learning: 54.34\% & Performance decreases under noisy conditions; MMSE used for enhancement; limited dataset size (322 sentences)\\

\hline
\cite{sc06} & 2010 & SVM with Large Acoustic Feature Set (4k LLDs + IGR-FS) & DES, EMO-DB, SUSAS & White Noise / Real-world Noise & DES: 74.5\% (clean), 54.9\% (-10dB SNR); EMO-DB: 87.5\% (clean), 71.1\% (-10dB SNR); SUSAS: 84.9\% & Feature selection improves accuracy; performance drops with severe noise; robust on real-world noisy SUSAS data \\

\hline
\cite{article1} & 2021 & Head Fusion (ACNN + Multi-head Attention) & IEMOCAP, RAVDESS & Real-world / Simulated ESC-50 noise & Weighted Acc. : 76.36\% , UnWeighted Acc. : 76.18\%  & Evaluated robustness under 50 types of noise, noise injection strategies, and varying SNRs \\

\hline
\cite{chak19} & 2019 & Front-end feature compensation (VTS, VTS-AM) \& TDNN-DAE denoising with feature selection & EmoDB, IEMOCAP & Babble, Factory, HF channel, F16, Volvo & Up to 79.71\% (EmoDB, TDNN-DAE, 20 dB Babble) & TDNN-DAE performs best in seen noise conditions; 10th-root compression with VTS-AM improves MFCC features; feature selection on non-MFCC HLDs provides consistent gain \\

\hline
\cite{10347397} & 2024 & Selective Acoustic Feature Enhancement (enhance only weak LLDs) & MSP-Podcast, IEMOCAP & Multi-noise, Real-world & MSP-Podcast: 17.7\% (arousal), 21.2\% (dominance), 3.3\% (valence) gain; IEMOCAP: 5.1\% (arousal), 7.9\% (dominance), 45.2\% (valence) gain & Feature-based enhancement outperforms signal-based; robust LLDs are left unchanged; works under matched, mismatched, and multi-SNR conditions \\

\hline
\cite{10018195}& 2023 & 
ASP-MTL (Adversarial Shared–Private Multi-Task Learning) & 
AVEC 2016, AFEW 6.0, IEMOCAP & 
Real-world noise + white/pink/babble /factory (various SNRs) & 
~10\% improvement in accuracy and F1 over prior works & 
Two-stage model; requires multi-task setup for noise and gender handling \\

\hline
\cite{8852707} & 2019 & LW-WPCC + SVM (with audio-visual fusion) & 
Berlin Emotional Corpus, IEMOCAP, NOISEX-92 & 
Realistic + synthetic noise (white, pink, car, restaurant, factory) across SNR 0–30dB & 72.12\% accuracy (Berlin); audio-visual fusion ~15\% higher accuracy under noise & Handcrafted feature extraction; multimodal fusion sensitive to lighting and face position \\

\hline
\cite{WI0165} & 2021 & FCNN with magnitude + Modified Group Delay (MGD) spectrograms; attention mechanism; noise-augmented training & Berlin Emotional Speech Database; RAVDESS & 9 indoor noise types + AWGN at SNRs 10,15,20,25,30,35 & Noise-augmented training + combined features improved noise robustness; attention mechanism helped focus on clean sections; Avg. 81\% under noise, 91\% clean  & Initial experiments on synthetic/indoor noise; real-world noise generalization not fully tested; class imbalance handled with class weights \\

\hline
\cite{103389} & 2021 & Multimodal AER (speech + text); MetricGAN+; Data augmentation; DNN classifier & MELD, IEMOCAP, Emoti-W & Babble, cafeteria, airplane noise; SNR –10 to 20 dB & Enhancement + augmentation improved cross-corpus AER by 17–44\% & Enhancement alone insufficient; cross-corpus still challenging \\

\hline
\cite{chen20} & 2023 & NRSER: SE (CMGAN) + SNR-level detection + waveform reconstitution; SSL-based emotion recognition & MSP-PODCAST, AudioSet, LibriSpeech & Mixed background noise, SNR 6–14 dB & Improved noise robustness; avoids recognizing background-only signals; best F1 for emotion and valence & Less effective on arousal and dominance; SE may reduce signal intensity \\

\hline
\cite{inpr1} & 2020 & LSTM-IRM; LSTM-PL-MTL with progressive learning + ISPP post-processing & IEMOCAP, CHEAVD 2.0, WSJ0 & Simulated and realistic background noise, SNR -5 to 15 dB & SE trained on emotional speech improves SER; T3-IRM + ISPP achieves best performance (44.13\% on CHEAVD noisy test) & SE models trained on non-emotional corpus may destroy emotion cues; overfitting possible with deeper LSTM layers \\

\hline
\cite{9597437} & 2023 & DLC / SLC, Mixture-of-Experts, DSN, DANN, MADDoG, HFFN & MSP-Improv, MOSEI, IEMOCAP & Environmental noise (natural, human, interior) with varied SNR profiles & DLC improves noise robustness; MoE-DLC boosts unimodal UAR; DSN shows strongest generalization to unseen noise &
SLC breaks temporal sequence; MoE weak on unseen noise; DLC applied only to acoustic stream \\

\hline
\cite{7523189}& 2016 &
Spectral Subtraction, Wiener Filter, MMSE + MFCC + HMM &
IEMOCAP & Real-world noise: airport, babble, train, car (SNR: 0–15 dB) &
Enhancement improves accuracy mainly for airport/babble; spectral subtraction best; limited gains for machine noise &
HMM-based; only 4 classes; enhancement fails on train/car noise \\

\hline
\end{tabular}
\end{table*}

\section{\textbf{Methodology}}
\label{sec:methodology}
\subsection{\textbf{Composite Corpus Construction}}
\subsubsection{RAVDESS Corpus}
The \textit{Ryerson Audio-Visual Database of Emotional Speech and Song} (RAVDESS) \cite{revdess} was selected for its high standard of perceptual validity and rigorous recording environment. Unlike crowd-sourced datasets, RAVDESS utilizes professional actors delivering lexically matched statements (e.g., ``Kids are talking by the door'') to ensure that emotional variance is not confounded by linguistic content. The dataset encompasses emotional categories such as: \textit{Happy, Sad, Angry, Neutral, etc.}. A key feature utilized in this study is the inclusion of two distinct emotional intensities—\textit{Normal} and \textit{Strong}—which challenges the model to detect subtle gradations of sentiment rather than just exaggerated caricatures.

\subsubsection{CREMA-D Corpus}
To ensure the model's generalization across diverse vocal textures, we incorporated the \textit{Crowd-sourced Emotional Multimodal Actors Dataset} (CREMA-D) \cite{6849440}. This dataset is pivotal for minimizing speaker-dependent information leakage, a common issue in SER where models overfit to specific voice actors rather than learning generalized emotional cues. CREMA-D provides a rich demographic cross-section, featuring actors from African American, Asian, Caucasian, and Hispanic backgrounds, with ages ranging from 20 to 74. The recording protocol involved actors speaking a selection of 12 standard sentences across four distinct intensity levels: \textit{Low, Medium, High,} and \textit{Unspecified}. This variation in intensity, combined with the ethnic diversity, forces the model to learn robust features that are invariant to accent and speaker identity.

\subsubsection{TESS Corpus}
To address the male-skewed distribution prevalent in standard SER benchmarks and ensure robust performance on female vocal characteristics, we integrated the \textit{Toronto Emotional Speech Set} (TESS) \cite{dupuis2011recognition}. This dataset is exclusively female, featuring high-fidelity recordings from two distinct subjects: a younger speaker (26 years) and a senior speaker (64 years). This age stratification provides critical data for analyzing emotional features across different stages of vocal maturity. It comprises 2,800 stimuli where 200 target words are embedded within the fixed carrier phrase, ``Say the word \underline{\hspace{0.5cm}}.'' The subjects, who are musically trained, portray seven emotional states with precise articulation. The inclusion of TESS is essential for balancing our composite dataset and preventing the model from developing a bias toward male frequency ranges.

\subsubsection{SAVEE Corpus}
The \textit{Surrey Audio-Visual Expressed Emotion} (SAVEE) \cite{jackson2014surrey} database was selected to provide a high-quality baseline for male emotional articulation, complementing the female-only TESS corpus. Recorded at the University of Surrey, this dataset features four native English male speakers (aged 27--31), all of whom were postgraduate researchers. The corpus consists of 480 utterances distributed across seven emotion categories: \textit{Anger, Disgust, Fear, Happiness, Sadness, Surprise,} and \textit{Neutral}. A defining characteristic of SAVEE is its linguistic rigor; the text material is derived from standard \textbf{TIMIT sentences}, which are phonetically balanced to ensure comprehensive coverage of phonemes. Each speaker recorded 15 sentences per emotion, including a mix of common, emotion-specific, and generic phrases. This structural consistency allows for precise analysis of how emotional inflection alters specific phonetic components in male speech.

The distribution of the aggregated corpus is detailed in Table~\ref{tab:dataset_dist}.

\vspace{-2mm}
\begin{table}[h]
\caption{Summary of the Composite Dataset Composition}
\begin{center}
\renewcommand{\arraystretch}{1.3}
\begin{tabular}{|l|c|p{0.8cm}|p{0.9cm}|p{3cm}|}
\hline
\textbf{Corpus} & \textbf{Spkrs} & \textbf{Gender} & \textbf{Samples} & \textbf{Key Characteristics} \\
\hline
CREMA-D & 91 & M/F & 7,442 & High ethnic diversity, varying vocal intensity \\
\hline
RAVDESS & 24 & M/F & 1,440 & Professional actors, neutral North American accent \\
\hline
TESS & 2 & F & 2,800 & Senior \& Young subjects; high-quality articulation \\
\hline
SAVEE & 4 & M & 480 & Male-specific; phone-tically balanced sentences \\
\hline
\textbf{Total} & \textbf{121} & \textbf{Mixed} & \textbf{$\sim$12,162} & \textbf{Unified Cross-Corpus} \\
\hline
\end{tabular}
\label{tab:dataset_dist}
\end{center}
\end{table}

\vspace{-3mm}

\begin{table}[htbp]
\caption{Detailed Class Distribution Across Source Corpora}
\begin{center}
\renewcommand{\arraystretch}{1.3}
\begin{tabular}{|>{\arraybackslash}m{1cm}|>{\centering\arraybackslash}m{1.3cm}|>{\centering\arraybackslash}m{1.4cm}|>{\centering\arraybackslash}m{0.8cm}|>{\centering\arraybackslash}m{0.9cm}|>{\centering\arraybackslash}m{0.8cm}|}
\hline
\textbf{Emotion Class} & \textbf{RAVDESS} & \textbf{CREMA-D} & \textbf{TESS} & \textbf{SAVEE} & \textbf{Total} \\
\hline
\textbf{Angry} & 192 & 1,271 & 400 & 60 & \textbf{1,923} \\
\hline
\textbf{Disgust} & 192 & 1,271 & 400 & 60 & \textbf{1,923} \\
\hline
\textbf{Fear} & 192 & 1,271 & 400 & 60 & \textbf{1,923} \\
\hline
\textbf{Happy} & 192 & 1,271 & 400 & 60 & \textbf{1,923} \\
\hline
\textbf{Neutral} & 288 & 1,087 & 400 & 120 & \textbf{1,895} \\
\hline
\textbf{Sad} & 192 & 1,271 & 400 & 60 & \textbf{1,923} \\
\hline
\textbf{Surprise} & 192 & --- & 400 & 60 & \textbf{652} \\
\hline
\textbf{TOTAL} & \textbf{1,440} & \textbf{7,442} & \textbf{2,800} & \textbf{480} & \textbf{12,162} \\
\hline
\end{tabular}
\label{tab:class_distribution}
\end{center}
\end{table}

\subsection{\textbf{Noise Injection Protocol}}
\subsubsection{Source of Non-Stationary Environmental Noise}
To simulate realistic, unstructured acoustic interference, we employed the \textit{South Asian Sound Dataset} (SAS-KIIT) \cite{10829485}. Unlike commonly used environmental noise benchmarks that predominantly contain sparse or regulated urban sounds, SAS-KIIT\footnote{SAS-KIIT Dataset: \url{https://sas-kiit.netlify.app/}} captures acoustically dense and highly non-stationary soundscapes characteristic of high-density South Asian urban environments. From this corpus, three representative noise categories were selected: \textit{Fish Market} (vocal babble), \textit{Railway Engine} (low-frequency mechanical rumble), and \textit{Children Classroom Noise} (impulsive, transient bursts).

\textbf{Dynamic Noise Injection Protocol:}
Rather than relying on pre-segmented or fixed-length noise clips, we adopted an on-the-fly mixing strategy to enhance distributional diversity. For each speech sample, the following procedure was applied:
\begin{enumerate}
    \item \textbf{Temporal Alignment:} A noise signal is randomly sampled and temporally matched to the speech duration. If the noise exceeds the speech length, it is truncated; if shorter, it is zero-padded (without looping). This choice preserves natural acoustic onsets and offsets while avoiding artificial periodicity introduced by noise repetition.
    
    \item \textbf{Stochastic Amplitude Scaling:} To avoid overfitting, noise intensity is scaled relative to the peak amplitude of the speech signal using a randomized factor. Let $s(t)$ denote the clean speech and $n(t)$ the selected noise. The normalized noise $\tilde{n}(t)$ is defined as:
    \begin{equation}
        \tilde{n}(t) = \frac{n(t)}{\max |n(t)|}
    \end{equation}
    The noise amplitude is then computed as:
    \begin{equation}
        A_{noise} = \lambda \cdot \max |s(t)|, \quad \lambda \sim U(0, 0.5, 0.75)
    \end{equation}
    and the corrupted signal is obtained as:
    \begin{equation}
        x(t) = s(t) + A_{noise} \cdot \tilde{n}(t)
    \end{equation}
    This stochastic scaling exposes the model to a continuous spectrum of interference levels, promoting robustness against varying real-world noise intensities.
\end{enumerate}

\subsubsection{Stationary Noise Augmentation Profiles}
To evaluate model robustness under controlled spectral conditions, we introduced three canonical stationary noise types: \textit{White}, \textit{Pink}, and \textit{Brown} (Brownian) noise. These profiles are mathematically distinguished by their Power Spectral Density (PSD), governed by the spectral decay law:
\begin{equation}
    S(f) \propto \frac{1}{f^\beta}
\end{equation}
where $\beta$ determines the spectral slope or ``color" of the noise.

\paragraph{White Noise ($\beta = 0$)}
White noise exhibits a flat PSD, distributing equal energy per Hertz across the bandwidth. It models broadband thermal noise and is generated by sampling from a standard normal distribution:
\begin{equation}
    x_{\text{white}}[n] = \mathcal{N}(0, \sigma^2)
\end{equation}
where $\mathcal{N}$ represents a Gaussian distribution with zero mean and variance $\sigma^2$.

\paragraph{Pink Noise ($\beta = 1$)}
Pink noise (flicker noise) is characterized by a $1/f$ spectral decay, meaning it carries equal energy per octave. This balance aligns closely with human auditory perception. It is approximated by filtering white noise through a transfer function $H(z)$ with magnitude response:
\begin{equation}
    |H(f)| \propto \frac{1}{\sqrt{f}}
\end{equation}
This filtering ensures the signal energy decreases by 3 dB per octave, simulating natural ambient background shifts.

\paragraph{Brownian Noise ($\beta = 2$)}
Brown noise (red noise) follows a $1/f^2$ profile, resulting in dominant low-frequency energy. It is generated via the cumulative integration of a white noise sequence, effectively simulating a discrete-time random walk:
\begin{equation}
    x_{\text{brown}}[n] = \sum_{k=0}^{n} x_{\text{white}}[k] = x_{\text{brown}}[n-1] + x_{\text{white}}[n]
\end{equation}
This formulation mirrors Brownian motion, effectively modeling heavy mechanical rumbles or distant traffic hums.

\paragraph{Discrete Mixing Protocol}
Unlike the continuous scaling used for non-stationary noise, these stationary profiles were injected at fixed discrete intervals to analyze specific failure points. The augmented signal $y[n]$ is constructed as:
\begin{equation}
    y[n] = x_{clean}[n] + \alpha \cdot \left( \frac{v[n]}{\max|v[n]|} \right)
\end{equation}
where the mixing coefficient $\alpha$ is selected from a discrete set of intensity ratios:
\begin{equation}
    \alpha \in \{0.25, 0.50, 0.75\}
\end{equation}
This creates a graded stress-test environment, corresponding to interference levels of 25\%, 50\%, and 75\% relative to the peak speech amplitude.

\subsection{\textbf{Data Augmentation Strategy}}
\label{subsec:augmentation}
To mitigate data scarcity and improve robustness to real-world acoustic variability, we employed a compound augmentation pipeline that expands the training set by a factor of four ($4\times$) via a \textit{Quadruplet Generation Protocol}.

\begin{enumerate}
    \item \textbf{Stochastic Environmental Noise Injection:} 
    Each clean speech signal was corrupted using non-stationary environmental noise characterized by broadband mechanical and crowd-like spectral components. Rather than enforcing fixed Signal-to-Noise Ratios (SNRs), we adopted an amplitude-based mixing strategy, where the noise signal is scaled relative to the peak speech amplitude:
    \begin{equation}
        x_{\text{noise}}(t) = x(t) + \lambda \cdot n(t), \quad 0 < \lambda < 0.75,
    \end{equation}
    allowing the model to encounter a continuum of interference intensities without overfitting to discrete SNR values.

    \item \textbf{Pitch Shifting:}
    To simulate speaker-dependent pitch variability and emotional prosody shifts, pitch modulation was applied without altering the temporal duration. Given a pitch shift of $n$ semitones, the operation can be expressed in the frequency domain as:
    \begin{equation}
        P(x(t)) = \mathcal{F}^{-1} \left\{ X \left( f \cdot 2^{n/12} \right) \right\},
    \end{equation}
    where $n = 0.7$ semitones in our experiments, effectively generating virtual speaker variants.
\end{enumerate}

\subsubsection{Quadruplet Generation Protocol}
For each input waveform $x(t)$, four training instances are generated and vertically stacked:
\begin{itemize}
    \item \textbf{Original:} Clean speech ($x$)
    \item \textbf{Noisy:} Environmental noise injection ($x + n$)
    \item \textbf{Pitched:} Pitch-shifted speech ($P(x)$)
    \item \textbf{Pitched \& Noisy:} Combined augmentation ($P(x) + n$)
\end{itemize}
This compound strategy enforces invariance to both harmonic perturbations (speaker variability) and spectral degradation (environmental interference).

\subsection{\textbf{Acoustic Feature Extraction}}
\label{subsec:feature_extraction}
Following augmentation, a short-time signal processing pipeline was employed to extract emotionally salient acoustic descriptors. All features were computed using a frame length of $N=2048$ samples and a hop size of $H=512$ samples.

For each audio instance, three complementary feature sets were computed and concatenated to form the final representation:
\begin{equation}
    \mathbf{f} = \left[ \mathbf{zcr} \;\|\, \mathbf{rmse} \;\|\, \mathbf{mfcc} \right].
\end{equation}

\subsubsection{Zero-Crossing Rate (ZCR)}
ZCR quantifies the rate of signal sign changes and serves as an indicator of signal noisiness and voicing:
\begin{equation}
ZCR_m = \frac{1}{2(N-1)} \sum_{n=1}^{N-1} |\text{sgn}(x[n]) - \text{sgn}(x[n-1])|.
\end{equation}

\subsubsection{Root Mean Square Energy (RMSE)}
To capture short-term intensity dynamics associated with emotional arousal, the RMS energy is computed as:
\begin{equation}
RMSE_m = \sqrt{\frac{1}{N} \sum_{n=0}^{N-1} |x[n]|^2}.
\end{equation}

\begin{figure*}[ht!]
    \centering
    \begin{minipage}[t]{0.99\textwidth}
        \centering
        \includegraphics[width=18.2cm, height=9cm]{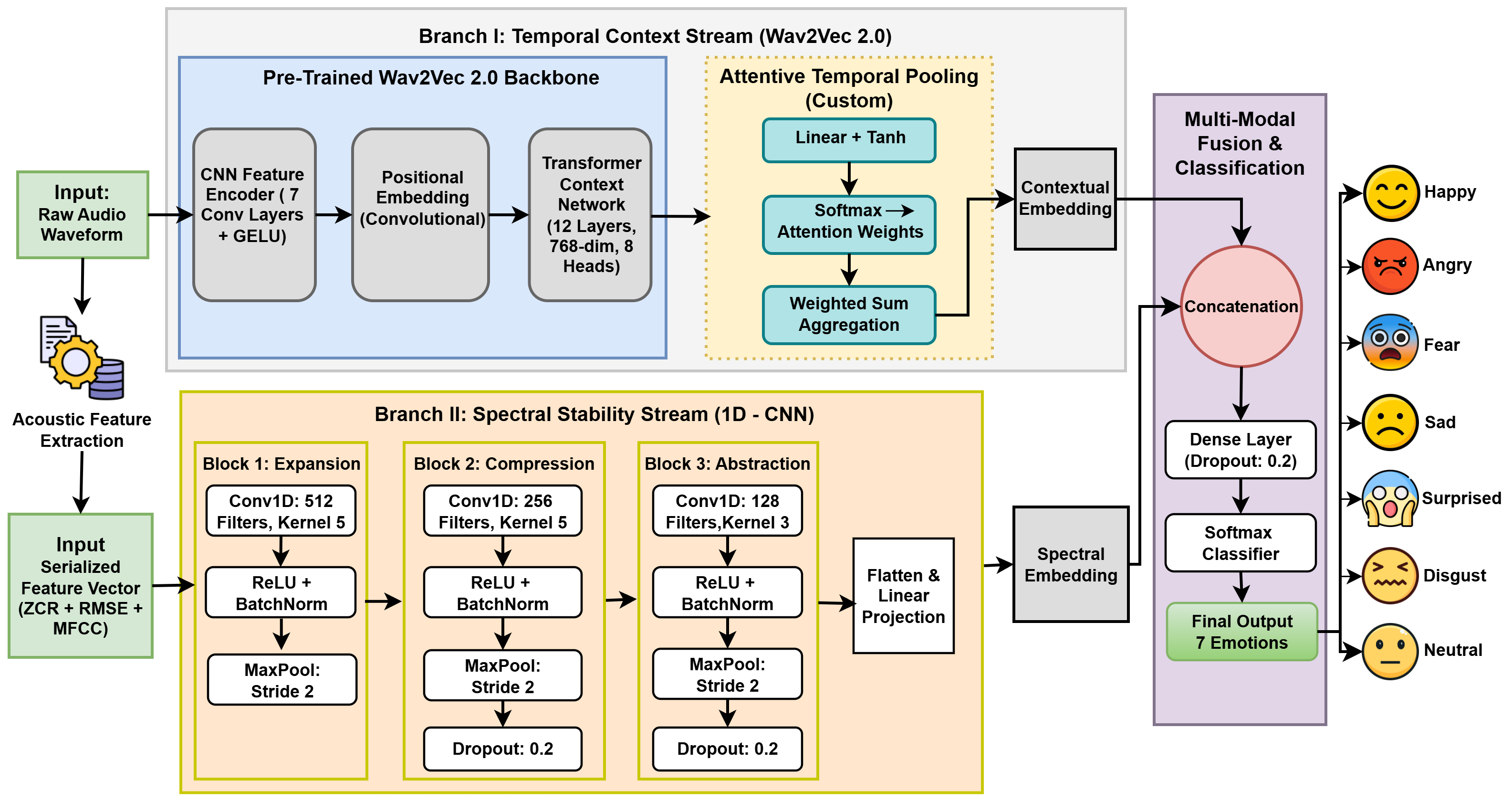}
        \caption{The proposed architecture combining a pre-trained Transformer backbone with a dedicated CNN feature encoder. Branch I utilizes Wav2Vec with a learnable attention mechanism to derive contextual embeddings, while Branch II extracts robust spectral embeddings from MFCC, RMSE, and ZCR inputs.}
        \label{fig:model_architecture}
    \end{minipage}\hfill
\end{figure*}

\subsubsection{Mel-Frequency Cepstral Coefficients (MFCCs)}
MFCCs capture the spectral envelope of speech signals on a perceptually motivated Mel scale. The extraction process consists of the following steps:

\begin{enumerate}
    \item \textbf{Short-Time Fourier Transform (STFT):} 
    For each frame $m$, the windowed STFT is computed as:
    \begin{equation}
    X_m(k) = \sum_{n=0}^{N-1} x(n + mH)\, w(n)\, e^{-j2\pi kn/N},
    \end{equation}
    where $w(n)$ is the analysis window, $N$ is the frame length, and $H$ is the hop size.

    \item \textbf{Power Spectrum:}
    \begin{equation}
    P_m(k) = |X_m(k)|^2.
    \end{equation}

    \item \textbf{Mel Filterbank Energy:}
    The power spectrum is projected onto $M$ Mel-scaled triangular filters $M_m(k)$:
    \begin{equation}
    E_m = \sum_{k=1}^{K} P_m(k)\, M_m(k),
    \end{equation}
    where $K$ denotes the number of frequency bins.

    \item \textbf{Logarithmic Compression:}
    \begin{equation}
    \hat{E}_m = \log(E_m).
    \end{equation}

    \item \textbf{Discrete Cosine Transform (DCT):}
    Finally, cepstral coefficients are obtained via the DCT:
    \begin{equation}
    \text{MFCC}(c) = \sum_{m=1}^{M} \hat{E}_m
    \cos\left[ \frac{\pi c}{M} (m - 0.5) \right],
    \end{equation}
    where $c$ indexes the cepstral coefficients.
\end{enumerate}

\subsection{\textbf{Multi-Stream Spectral-Temporal Fusion Network}}
We propose a dual-branch architecture designed to mitigate the trade-off between semantic contextualization and acoustic stability. Let $\mathcal{X}_{raw} \in \mathbb{R}^T$ denote the raw audio waveform and $\mathcal{X}_{spec} \in \mathbb{R}^{L}$ denote the serialized handcrafted feature vector.

\subsubsection{Branch I: Temporal Context Stream (Wav2Vec 2.0)}
The first branch leverages the Wav2Vec 2.0 backbone. The raw waveform is processed by the feature encoder $\Phi_{enc}$ and Transformer context network $\Phi_{tr}$, yielding hidden states $H = \{h_1, \dots, h_T\}$.

To aggregate these states, we implement an \textbf{Attentive Temporal Pooling} head. The attention scores are computed as:
\begin{equation}
    e_t = \mathbf{w}^T \tanh(\mathbf{W}_a h_t + \mathbf{b}_a)
\end{equation}

\begin{equation}
    \alpha_t = \frac{\exp(e_t)}{\sum_{k=1}^{T} \exp(e_k)}
\end{equation}
where $\mathbf{W}_a$ and $\mathbf{w}$ are learnable weights (bias disabled for $\mathbf{w}$). The final context vector is the weighted sum $v_{ctx} = \sum \alpha_t h_t$.

\subsubsection{Branch II: Spectral Stability Stream (1D-CNN)}
The second branch processes the feature vector $\mathcal{X}_{spec}$ (comprising ZCR, RMSE, and MFCCs) via a custom \textbf{1D-CNN}. The network follows a ``Compression-Abstraction" hierarchy defined by three sequential blocks:

\begin{enumerate}
    \item \textbf{Expansion Block:} Conv1D ($k=5$, 512 filters) $\to$ BatchNorm $\to$ ReLU $\to$ MaxPool ($k=5$, stride=2).
    \item \textbf{Compression Block:} Conv1D ($k=5$, 256 filters) $\to$ BatchNorm $\to$ ReLU $\to$ MaxPool ($k=5$, stride=2) $\to$ Dropout ($p=0.2$).
    \item \textbf{Abstraction Block:} Conv1D ($k=3$, 128 filters) $\to$ BatchNorm $\to$ ReLU $\to$ MaxPool ($k=3$, stride=2) $\to$ Dropout ($p=0.2$).
\end{enumerate}

The output is flattened and passed through a dense projection head (Linear $\to$ ReLU $\to$ BatchNorm) to yield the spectral embedding $v_{spec} \in \mathbb{R}^{512}$.

\subsubsection{Multi-Modal Fusion and Classification}
The semantic view ($v_{ctx}$) and physical view ($v_{spec}$) are fused via concatenation:
\begin{equation}
    v_{fusion} = v_{ctx} \oplus v_{spec}
\end{equation}
The joint representation is passed through a classification MLP:
\begin{equation}
    \hat{y} = \text{Softmax}\left( \mathbf{W}_{out} \cdot \delta_{0.5} \left( \sigma(\mathbf{W}_{d} v_{fusion} + \mathbf{b}_d) \right) \right)
\end{equation}
\vspace{1mm}
where $\sigma$ is the ReLU activation and $\delta_{0.5}$ denotes Dropout with $p=0.5$. This higher dropout rate ensures the model does not over-rely on the dominant modality.

The final output of the architecture consists of a classification into seven distinct emotional classes: Happy, Angry, Fear, Sad, Surprised, Disgust, and Neutral.

\subsection{\textbf{Explainable AI (XAI) Framework}}
To ensure the proposed Fusion Network relies on acoustically relevant features rather than spurious correlations (e.g., background noise), we integrated a post-hoc interpretability framework comprising two distinct approaches: SHAP \cite{NIPS2017_8a20a862} for feature attribution and Score-CAM \cite{wang2020score} for visual localization.

\subsubsection{SHAP (SHapley Additive exPlanations)}
SHAP is a game-theoretic approach that assigns each feature an importance value representing its contribution to the model's output. Unlike local surrogate models, SHAP guarantees consistency and local accuracy by computing the Shapley value, which is the average marginal contribution of a feature value across all possible coalitions.

Let $f$ be the trained fusion model and $x$ be the input instance. The SHAP value $\phi_i$ for a specific feature $i$ (where $i$ corresponds to a specific time-step or spectral coefficient in our flattened feature vector) is defined as:

\begin{equation}
    \phi_i(f, x) = \sum_{S \subseteq F \setminus \{i\}} \frac{|S|! (|F| - |S| - 1)!}{|F|!} \left[ f_x(S \cup \{i\}) - f_x(S) \right]
\end{equation}

where:
\begin{itemize}
    \item $F$ is the set of all input features.
    \item $S$ is a subset of features excluding $i$.
    \item $f_x(S)$ is the prediction of the model when only features in subset $S$ are present (marginalized over the rest).
\end{itemize}
In the context of our SER task, a positive $\phi_i$ indicates that the $i$-th time-step pushes the prediction toward the target emotion (e.g., `Fear'), while a negative value suppresses it. Mathematically, SHAP decomposes the model's prediction $f(x)$ into a sum of feature attribution values:

\begin{equation}
    f(x) = \phi_0 + \sum_{i=1}^{M} \phi_i
\end{equation}

\noindent where $\phi_0$ is the base value (average model output) and $\phi_i$ represents the contribution of the $i$-th feature. In the context of our SER task:
\begin{itemize}
    \item A \textbf{positive} $\phi_i$ indicates that the specific acoustic segment pushes the prediction \textit{toward} the target emotion (e.g., `Fear').
    \item A \textbf{negative} $\phi_i$ implies that the segment \textit{suppresses} the prediction, pushing it towards a different class.
\end{itemize}

\vspace{2mm}

\subsubsection{Score-CAM (Score-Weighted Class Activation Mapping)}
To visualize the temporal focus of the convolutional branch (Spectral Stability Stream), we employed Score-CAM. Unlike gradient-based methods (e.g., Grad-CAM) which can suffer from gradient saturation or instability in noisy environments, Score-CAM is gradient-free. It determines the importance of feature maps by measuring the \textit{change in confidence} when the input is masked by that specific map.

Let $A^k$ represent the $k$-th feature map in the final convolutional layer of Branch II. The class-discriminative saliency map $L_{Score-CAM}^c$ for a target emotion $c$ is computed in two phases:

\textbf{Phase 1: Importance Weight Computation.} 
We first upsample $A^k$ to match the input size and normalize it to $s(A^k) \in [0, 1]$. This map acts as a mask on the original input $X$ to obtain the masked input $M^k = X \circ s(A^k)$. The importance weight $\alpha_k^c$ is defined as the increase in model confidence for class $c$ relative to a baseline:
\begin{equation}
    \alpha_k^c = F^c(X \circ s(A^k)) - F^c(X_b)
\end{equation}
where $F^c(\cdot)$ is the model's output score for class $c$ and $X_b$ is a baseline input (typically zero).

\textbf{Phase 2: Linear Combination.}
The final heatmap is generated by a linear combination of feature maps weighted by their importance scores, followed by a ReLU activation to isolate features that have a positive influence on the target class:
\begin{equation}
    L_{Score-CAM}^c = \text{ReLU}\left( \sum_{k} \alpha_k^c A^k \right)
\end{equation}

The complete procedural flow for generating these class-discriminative saliency maps is formalized in \textbf{Algorithm \ref{alg:score_cam}}.

\begin{algorithm}[h]
\SetAlgoLined
\KwIn{Input Spectrogram $X$, Trained Model $F(\cdot)$, Target Class $c$}
\KwOut{Saliency Map $L_{Score-CAM}^c$}

\textbf{Initialization:}\\
$A \leftarrow$ GetFeatureMaps($F$, Layer='last\_conv')\;
$N \leftarrow$ Number of filters in $A$\;
$S \leftarrow$ ZeroArray(Shape of $X$)\;

\textbf{Compute Importance Weights:}\\
\For{$k \leftarrow 1$ \KwTo $N$}{
    $M^k \leftarrow$ Upsample($A^k$) to size of $X$\;
    $M^k_{norm} \leftarrow$ Normalize($M^k$) to $[0,1]$\;
    $X_{masked} \leftarrow X \circ M^k_{norm}$ \tcp*{Apply Mask}
    $Score_k \leftarrow F^c(X_{masked}) - F^c(X_{baseline})$\;
    $\alpha_k^c \leftarrow Score_k$\;
}

\textbf{Generate Heatmap:}\\
$L_{raw} \leftarrow \sum_{k=1}^{N} \alpha_k^c \cdot A^k$\;
$L_{Score-CAM}^c \leftarrow \text{ReLU}(L_{raw})$\;

\Return $L_{Score-CAM}^c$
\caption{Score-CAM Saliency Map Generation}
\label{alg:score_cam}
\end{algorithm}

This methodology ensures that the generated heatmaps highlight regions that are \textit{causally} linked to the prediction, providing a robust visualization of the model's attention mechanism.

\section{\textbf{Experimental Results and Analysis}}
\label{sec:results_analysis}
\subsection{\textbf{Performance Evaluation under Clean Conditions}}
To establish a performance baseline, we first evaluated the proposed Fusion Network on the clean, unaugmented test sets of each individual corpus. Table \ref{tab:clean_results} summarizes the classification metrics (Accuracy, Precision, Recall, and F1-Score) achieved when the model is trained and tested on the respective datasets without noise injection.

\begin{table}[h!]
\caption{Performance Metrics on Individual Clean Datasets}
\renewcommand{\arraystretch}{1.4}
\begin{center}
\begin{tabular}{|l|c|c|c|c|}
\hline
\textbf{Dataset} & \textbf{Acc. (\%)} & \textbf{Precision} & \textbf{Recall} & \textbf{F1-Score} \\
\hline
\textbf{TESS} & \textbf{99.78} & 0.9932 & 0.9945 & \textbf{0.9959} \\
\hline
\textbf{CREMA-D} & 97.93 & 0.9797 & 0.9794 & 0.9795 \\
\hline
\textbf{RAVDESS} & 91.28 & 0.9259 & 0.9175 & 0.9137 \\
\hline
\textbf{SAVEE} & 88.32 & 0.8942 & 0.8865 & 0.8875 \\
\hline
\end{tabular}
\label{tab:clean_results}
\end{center}
\end{table}

\textbf{Analysis of Clean Performance:} 
The model demonstrates exceptional performance on the \textbf{TESS} dataset, achieving near-perfect accuracy (99.78\%). This is attributed to the high-fidelity recording environment and the distinct, exaggerated emotional articulation of the actors, which minimizes intra-class variance. In contrast, performance on \textbf{SAVEE} is slightly lower (88.32\%). This dip is expected due to the dataset's limited size (only 480 utterances), which restricts the model's ability to generalize complex feature representations. \textbf{CREMA-D} yields a robust 97.93\%, validating the model's capacity to handle diverse speaker demographics without overfitting. The slight drop in \textbf{RAVDESS} (91.28\%) suggests that the ``Normal" vs. ``Strong" intensity variations in this dataset present a more challenging classification boundary than the binary emotions in TESS.

\begin{table*}[ht!]
\caption{Model Performance (Mean $\pm$ SD) under Stationary and Non-Stationary Noise Conditions}
\begin{center}
\renewcommand{\arraystretch}{1.5} 
\begin{tabular}{|m{2.5cm}|m{2.0cm}|m{2.5cm}|m{2.5cm}|m{2.5cm}|m{2.5cm}|}
\hline
\centering\textbf{Noise Source} & \centering\textbf{Intensity ($\lambda$)} & \centering\textbf{Accuracy (\%)} & \centering\textbf{Precision} & \centering\textbf{Recall} & \centering\textbf{F1-Score} \tabularnewline
\hline
\hline
\multicolumn{6}{|c|}{\textit{\textbf{Stationary Interferences}}} \tabularnewline
\hline
\multirow{3}{*}{\centering\textbf{White Noise}} 
 & \centering Low (0.25)    & \centering 98.29 $\pm$ 0.075 & \centering 0.9851 $\pm$ 0.022 & \centering 0.9812 $\pm$ 0.032 & \centering 0.9819 $\pm$ 0.042 \tabularnewline \cline{2-6}
 & \centering Medium (0.50) & \centering 95.67 $\pm$ 0.092 & \centering 0.9543 $\pm$ 0.073 & \centering 0.9523 $\pm$ 0.043 & \centering 0.9532 $\pm$ 0.068 \tabularnewline \cline{2-6}
 & \centering High (0.75)   & \centering 93.17 $\pm$ 0.105 & \centering 0.9326 $\pm$ 0.094 & \centering 0.9306 $\pm$ 0.074 & \centering 0.9325 $\pm$ 0.084 \tabularnewline
\hline
\multirow{3}{*}{\centering\textbf{Pink Noise}} 
 & \centering Low (0.25)    & \centering 96.33 $\pm$ 0.21 & \centering 0.9608 $\pm$ 0.05 & \centering 0.9656 $\pm$ 0.04 & \centering 0.9597 $\pm$ 0.04 \tabularnewline \cline{2-6}
 & \centering Medium (0.50) & \centering 93.21 $\pm$ 0.30 & \centering 0.9356 $\pm$ 0.06 & \centering 0.9267 $\pm$ 0.05 & \centering 0.9321 $\pm$ 0.05 \tabularnewline \cline{2-6}
 & \centering High (0.75)   & \centering 90.37 $\pm$ 0.41 & \centering 0.9093 $\pm$ 0.08 & \centering 0.8994 $\pm$ 0.07 & \centering 0.9037 $\pm$ 0.07 \tabularnewline
\hline
\multirow{3}{*}{\centering\textbf{Brown Noise}} 
 & \centering Low (0.25)    & \centering 98.68 $\pm$ 0.12 & \centering 0.9856 $\pm$ 0.01 & \centering 0.9821 $\pm$ 0.02 & \centering 0.9843 $\pm$ 0.01 \tabularnewline \cline{2-6}
 & \centering Medium (0.50) & \centering 96.09 $\pm$ 0.19 & \centering 0.9578 $\pm$ 0.03 & \centering 0.9573 $\pm$ 0.03 & \centering 0.9569 $\pm$ 0.02 \tabularnewline \cline{2-6}
 & \centering High (0.75)   & \centering 95.78 $\pm$ 0.25 & \centering 0.9589 $\pm$ 0.04 & \centering 0.9512 $\pm$ 0.04 & \centering 0.9518 $\pm$ 0.03 \tabularnewline
\hline
\multicolumn{6}{|c|}{\textit{\textbf{Non-Stationary Interferences}}} \tabularnewline
\hline
\multirow{3}{*}{\centering\textbf{Railway Engine}} 
 & \centering Low (0.25)    & \centering 97.28 $\pm$ 0.31 & \centering 0.9721 $\pm$ 0.04 & \centering 0.9716 $\pm$ 0.03 & \centering 0.9716 $\pm$ 0.03 \tabularnewline \cline{2-6}
 & \centering Medium (0.50) & \centering 95.09 $\pm$ 0.45 & \centering 0.9504 $\pm$ 0.06 & \centering 0.9578 $\pm$ 0.05 & \centering 0.9523 $\pm$ 0.05 \tabularnewline \cline{2-6}
 & \centering High (0.75)   & \centering 93.85 $\pm$ 0.52 & \centering 0.9401 $\pm$ 0.08 & \centering 0.9317 $\pm$ 0.07 & \centering 0.9392 $\pm$ 0.07 \tabularnewline
\hline
\multirow{3}{*}{\centering\textbf{Children Playing}} 
 & \centering Low (0.25)    & \centering 95.14 $\pm$ 0.35 & \centering 0.9472 $\pm$ 0.05 & \centering 0.9412 $\pm$ 0.06 & \centering 0.9492 $\pm$ 0.05 \tabularnewline \cline{2-6}
 & \centering Medium (0.50) & \centering 90.92 $\pm$ 0.61 & \centering 0.9031 $\pm$ 0.09 & \centering 0.8929 $\pm$ 0.08 & \centering 0.9047 $\pm$ 0.08 \tabularnewline \cline{2-6}
 & \centering High (0.75)   & \centering 85.43 $\pm$ 0.88 & \centering 0.8515 $\pm$ 0.12 & \centering 0.8489 $\pm$ 0.11 & \centering 0.8532 $\pm$ 0.11 \tabularnewline
\hline
\multirow{3}{*}{\centering\textbf{Fish Market}} 
 & \centering Low (0.25)    & \centering 94.13 $\pm$ 0.42 & \centering 0.9423 $\pm$ 0.06 & \centering 0.9349 $\pm$ 0.05 & \centering 0.9421 $\pm$ 0.05 \tabularnewline \cline{2-6}
 & \centering Medium (0.50) & \centering 90.34 $\pm$ 0.65 & \centering 0.8902 $\pm$ 0.10 & \centering 0.9021 $\pm$ 0.09 & \centering 0.8921 $\pm$ 0.10 \tabularnewline \cline{2-6}
 & \centering High (0.75)   & \centering 82.79 $\pm$ 1.12 & \centering 0.8234 $\pm$ 0.15 & \centering 0.8139 $\pm$ 0.14 & \centering 0.8312 $\pm$ 0.14 \tabularnewline
\hline
\end{tabular}
\label{tab:final_results}
\end{center}
\end{table*}

\begin{table*}[ht]
    \centering
    \caption{\textbf{Ablation Study: Component Contributions, F1 Scores, and Performance Gaps.}}
    \label{tab:ablation_study}
    \renewcommand{\arraystretch}{1.5} 
    \footnotesize 
    \begin{tabular}{|m{3.4cm}|m{1cm}|m{1.1cm}|m{1.1cm}|m{1.1cm}|m{1.1cm}|m{1.1cm}|m{1.1cm}|m{1.1cm}|m{1.1cm}|}
        \hline
        \textbf{Model Configuration} & 
        \centering\textbf{Params (M)} & 
        \centering\textbf{Clean Acc (\%)} & 
        \centering\textbf{Clean F1} & 
        \centering\textbf{Clean Drop} & 
        \centering\textbf{Inference Speed (Samp/s)} & 
        \centering\textbf{Noisy Acc (\%)} & 
        \centering\textbf{Noisy F1} & 
        \centering\textbf{Noisy Drop} & 
        \centering\textbf{Inference Speed (Samp/s)} \tabularnewline
        \hline
        \hline
        
        \textbf{A.} Spectral Branch Only \newline (1D-CNN) & 
        \centering 11.9 & \centering 93.94 & \centering 0.9359 & \centering \textcolor{red}{$\downarrow$ 4.97\%} & \centering 135 & \centering 91.73 & \centering 0.9184 & \centering \textcolor{red}{$\downarrow$ 5.55\%} & \centering 137 \tabularnewline
        \hline
        
        \textbf{B.} Temporal Branch Only \newline (Wav2Vec 2.0) & 
        \centering 94.5 & \centering 89.25 & \centering 0.8864 & \centering \textcolor{red}{$\downarrow$ 9.66\%} & \centering 80.0 & \centering 84.93 & \centering 0.8473 & \centering \textcolor{red}{$\textbf{$\downarrow$ 12.35\%}$} & \centering 80 \tabularnewline
        \hline
        
        \textbf{C.} Hybrid w/ Simple Concat \newline (No Attentive Pooling) & 
        \centering 114.6 & \centering 95.57 & \centering 0.9559 & \centering \textcolor{red}{$\downarrow$ 3.34\%} & \centering 78.0 & \centering 94.73 & \centering 0.9462 & \centering \textcolor{red}{$\downarrow$ 2.55\%} & \centering 78 \tabularnewline
        \hline
        
        \textbf{D. Proposed Model} \newline (Hybrid + Attentive Pooling) & 
        \centering \textbf{115.3} & \centering \textbf{98.91} & \centering \textbf{0.9769} & \centering \textbf{--} & \centering 74.0 & \centering \textbf{97.28} & \centering \textbf{0.9705} & \centering \textbf{--} & \centering 75 \tabularnewline
        \hline
    \end{tabular}
\end{table*}

\subsection{\textbf{Robustness Analysis under Acoustic Interference}}
To strictly evaluate the efficacy of the proposed dual-branch fusion mechanism, we subjected the model to a rigorous stress test involving both stationary and non-stationary spectral degradation. \textbf{Table \ref{tab:final_results}} details the performance metrics across the discrete noise intensity levels formally defined in Eq. (9).

\subsubsection{Performance in Stationary Environments}
The model demonstrated exceptional stability against stationary perturbations. Under Brown noise, which mimics the low-frequency spectral decay of ambient urban hums, the system retained a near-optimal accuracy of \textbf{95.78\%} even at the highest intensity level. Similarly, for White noise, performance degraded marginally by only $\approx 5\%$ (from 98.29\% to 93.17\%) despite the aggressive broadband masking. This suggests that the Spectral Stability Stream (1D-CNN) successfully anchored the decision boundary using robust energy features (RMSE/ZCR) when the spectral envelope was uniformly corrupted.

\subsubsection{Resilience to Non-Stationary degradation}
Non-stationary environments, characterized by unpredictable transient bursts, posed a greater challenge. As shown in the lower section of \textbf{Table \ref{tab:final_results}}:
\begin{itemize}
    \item \textbf{Railway Noise:} The model maintained high robustness (93.85\% at high intensity), likely because the mechanical rumbles share spectral similarities with Brown noise, which the CNN branch handles effectively.
    \item \textbf{Crowd \& Babble Noise:} The most significant performance drop was observed in the \textit{Fish Market} condition, where accuracy declined to \textbf{82.79\%} at the maximum interference level. This is attributed to the ``Cocktail Party Effect," \cite{cherry1953some} where overlapping vocal babble in the background competes directly with the foreground speech in the Wav2Vec 2.0 context stream.
\end{itemize}
However, even in these extreme conditions, the model maintained an F1-score above \textbf{0.83}, validating that the fusion of semantic attention and physical spectral descriptors prevents catastrophic failure in chaotic acoustic scenes.

\subsection{\textbf{Ablation Study: Contribution of Network Components}}
To validate the architectural contribution of each component, we conducted a comprehensive ablation study summarized in \textbf{Table \ref{tab:ablation_study}}. The standalone Wav2Vec 2.0 backbone (Row B) exhibits a significant performance drop from 89.25\% to 84.93\% when exposed to noise, confirming the fragility of self-supervised features in non-stationary environments. In contrast, the lightweight 1D-CNN branch (Row A) demonstrates higher intrinsic robustness (91.73\% noisy accuracy), suggesting that low-level spectral stability features remain reliable even when high-level context is corrupted. While merging these streams via simple concatenation (Row C) yields a notable improvement to 94.73\%, the definitive performance leap is achieved only with our proposed \textbf{Attentive Temporal Pooling} (Row D). This configuration achieves a state-of-the-art noisy accuracy of \textbf{97.28\%}, representing a massive \textbf{+12.35\% gain} over the Wav2Vec baseline with only a marginal increase in inference latency, effectively validating the necessity of the dual-stream fusion strategy.


\begin{figure*}[ht]
    \centering
    \begin{minipage}[t]{0.99\textwidth}
        \centering
        \includegraphics[width=18cm, height=12cm]{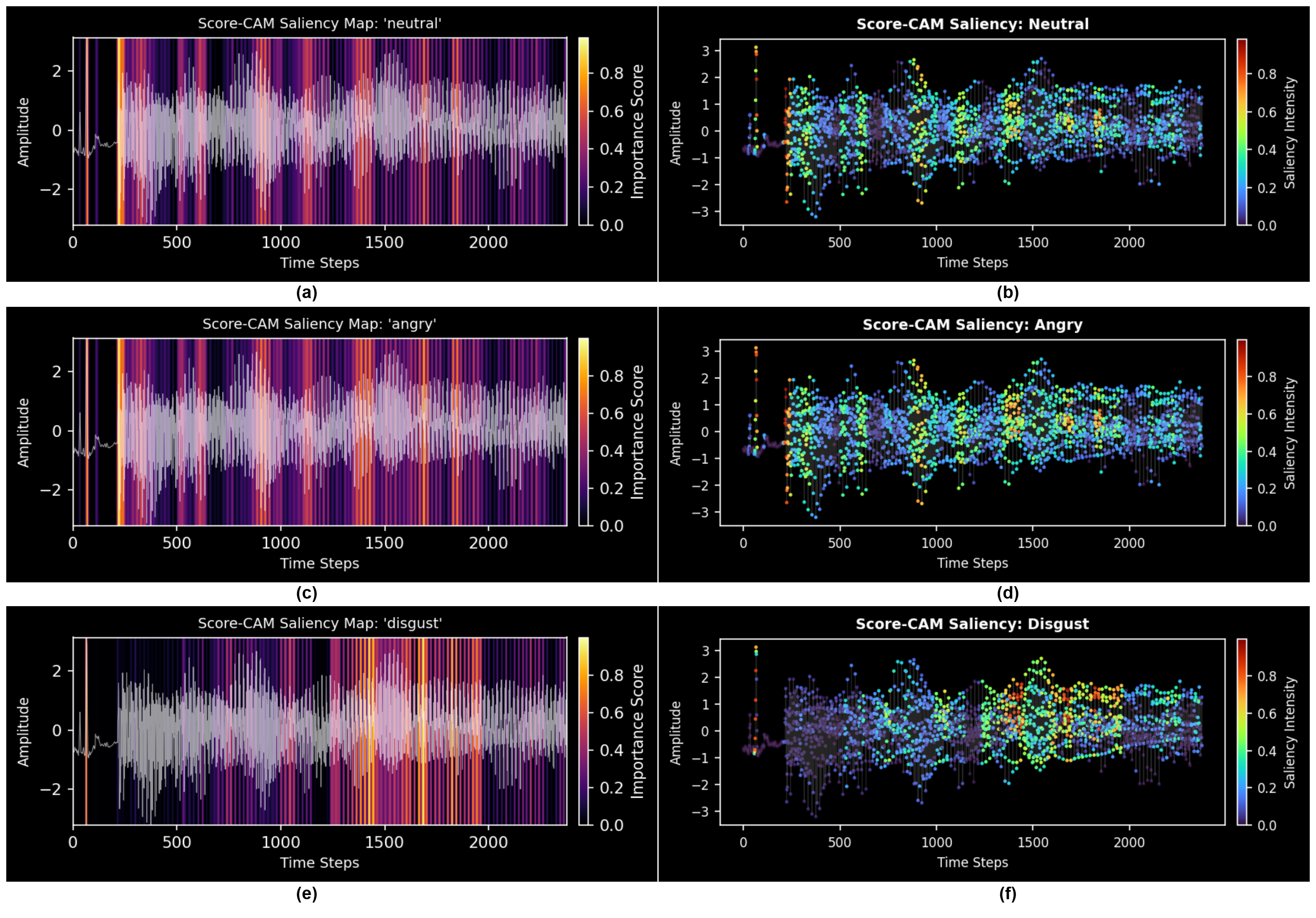}
        \caption{\textbf{Score-CAM Saliency Visualization.} The figure displays heatmaps for (a-b) Neutral, (c-d) Angry, and (e-f) Disgust emotions. In the left panels, the waveform is overlaid with a heatmap where \textbf{red/yellow regions} denote high importance (positive contribution to the prediction) and \textbf{blue/purple regions} denote low importance. The right panels show the corresponding discrete saliency intensity per time step.}
        \label{fig:saliency_maps}
    \end{minipage}\hfill
\end{figure*}

\section{\textbf{Explainability and Model Interpretability}}
\label{sec:xai}
While deep neural networks like Wav2Vec 2.0 and CNNs achieve state-of-the-art performance in Speech Emotion Recognition (SER), they are frequently criticized as ``black boxes" due to their opaque decision-making processes. In safety-critical applications, it is insufficient to know \textit{what} the model predicted; we must also understand \textit{why}.

\begin{figure*}[ht]
    \centering
    \begin{minipage}[t]{0.99\textwidth}
        \centering
        \includegraphics[width=18cm, height=13.3cm]{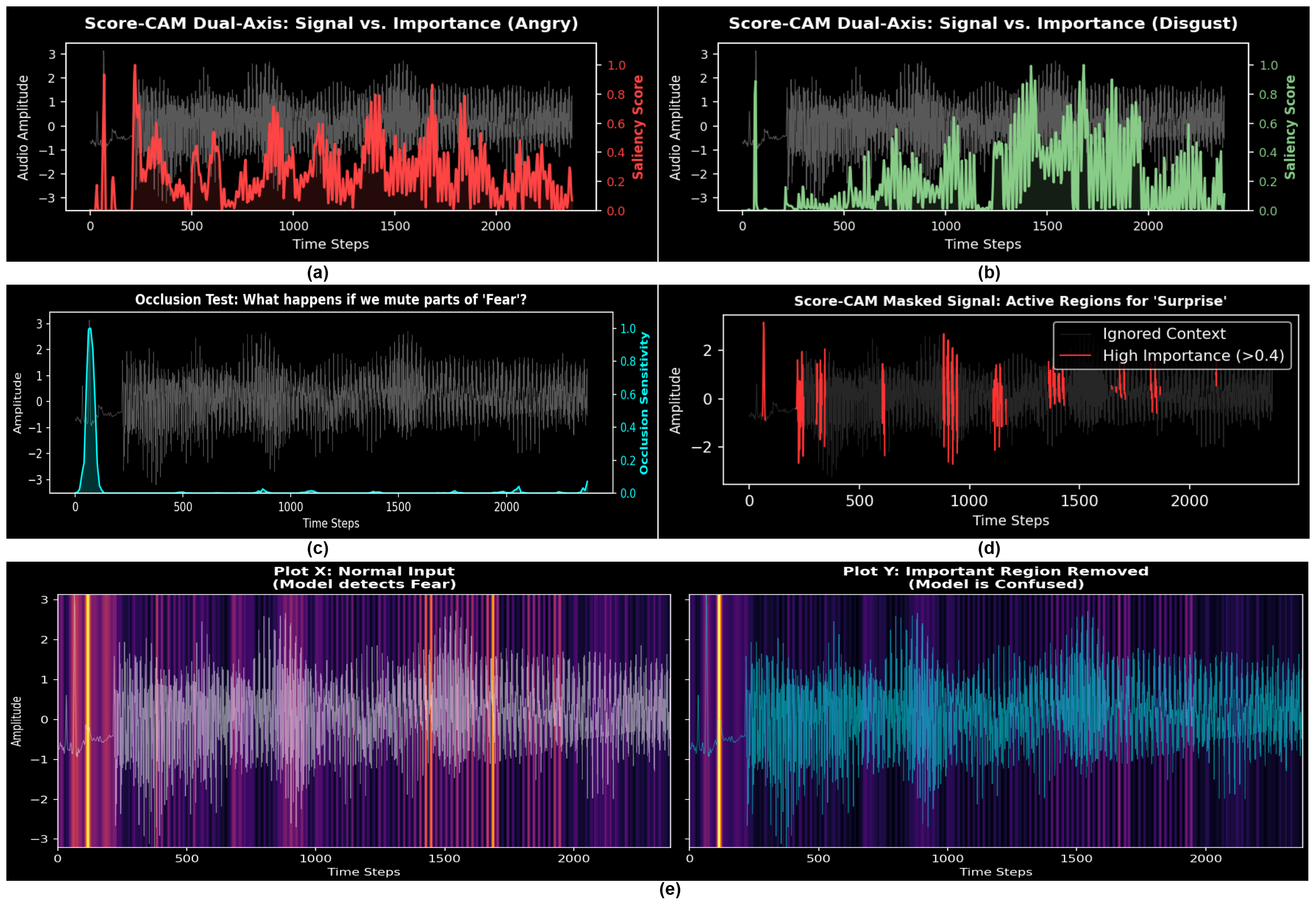}
        \caption{\textbf{Signal Alignment and Causal Perturbation Testing.} 
        \textbf{(a-b)} Dual-axis plots showing the alignment of importance scores with the raw waveform for `Angry' and `Disgust'. 
        \textbf{(c)} Occlusion sensitivity analysis for the 'Fear' class. 
        \textbf{(d)} Visualization of active vs. ignored regions for `Surprise'. 
        \textbf{(e)} Counterfactual perturbation test comparing the original signal (Plot X) against a masked signal where high-importance regions are removed (Plot Y).}
        \label{fig:perturbation_analysis}
    \end{minipage}\hfill
\end{figure*}

\subsection{\textbf{Score-Weighted Class Activation Mapping (Score-CAM)}}
\subsubsection{Class-Discriminative Saliency Analysis}
To validate the model's focus, we analyze the Score-CAM heatmaps presented in \textbf{Fig. \ref{fig:saliency_maps}}. The color gradient serves as a direct proxy for feature relevance: \textbf{red/yellow zones} indicate regions where the model attends most strongly to make a prediction, while \textbf{blue zones} represent suppressed or ignored data. As observed in the `Angry' samples (Fig. \ref{fig:saliency_maps}c), the model generates intense red bands that align perfectly with high-amplitude prosodic bursts, confirming its ability to leverage acoustic stress for arousal detection. Conversely, for `Neutral' speech (Fig. \ref{fig:saliency_maps}a), the saliency is sparsely distributed with no dominant red clusters, indicating that the model correctly interprets the absence of strong emotional transients as a defining characteristic of neutrality. This selective attention mechanism validates that the network is not overfitting to background silence but is instead grounding its decisions in linguistically significant audio segments.

\subsubsection{Causal Validation via Temporal Perturbation Analysis}
To ensure the network is not relying on spurious background correlations, we conducted a rigorous temporal perturbation analysis visualized in \textbf{Fig. \ref{fig:perturbation_analysis}}. 

First, the dual-axis alignment in \textbf{Fig. \ref{fig:perturbation_analysis}(a-b)} demonstrates that the saliency scores—shown as red and green traces for `Angry' and `Disgust' respectively—tightly align with the voiced, high-energy segments of the audio waveform (grey background). This confirms that the attention mechanism effectively suppresses non-informative silence. 

Second, the occlusion sensitivity test for `Fear' in \textbf{Fig. \ref{fig:perturbation_analysis}(c)} reveals that muting specific narrow time windows generates significant spikes in sensitivity (cyan curve), pinpointing the exact acoustic transients critical for decision-making. Similarly, for `Surprise' in \textbf{Fig. \ref{fig:perturbation_analysis}(d)}, the model isolates distinct prosodic onsets (red) while actively discarding the surrounding context. 

Finally, \textbf{Fig. \ref{fig:perturbation_analysis}(e)} provides definitive causal evidence: while the original signal (Plot X) is correctly classified, zeroing out \textit{only} the high-importance regions (Plot Y) causes the model's confidence to collapse. This proves that these specific prosodic features are causally responsible for the classification, rather than global environmental statistics.

\begin{figure*}[ht]
    \centering
    \begin{minipage}[t]{0.99\textwidth}
        \centering
        \includegraphics[width=17cm, height=14cm]{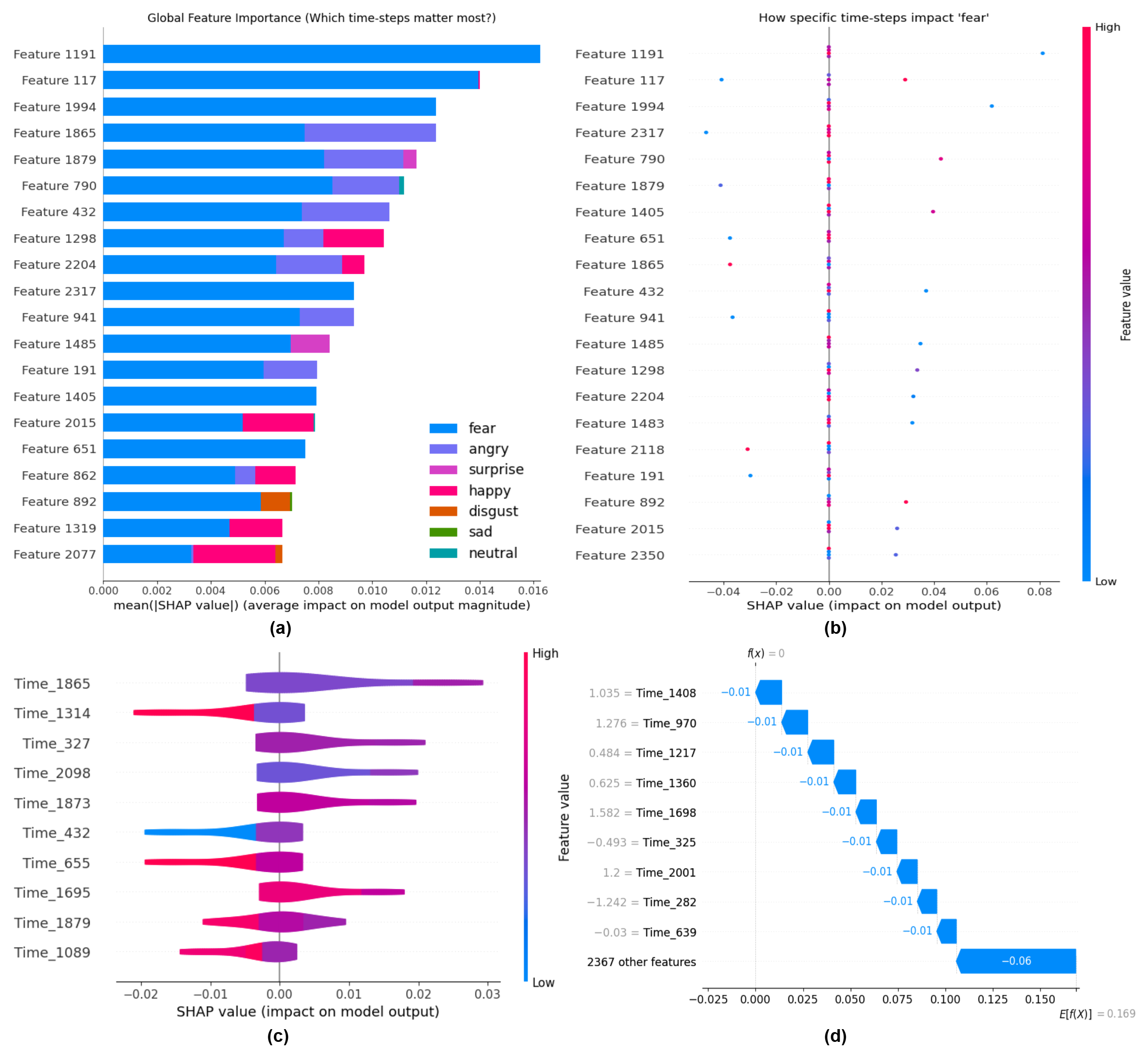}
        \caption{\textbf{SHAP Feature Contribution Analysis.} This multi-panel figure quantifies the impact of specific temporal features on the model's predictions. Panel \textbf{(a)} ranks the top features by global importance across all emotions. Panels \textbf{(b)} and \textbf{(c)} visualize the distribution of feature impacts specifically for the `fear' class. Panel \textbf{(d)} provides a local breakdown, showing how individual feature contributions accumulate to form a single prediction score.}
        \label{fig:shap_analysis}
    \end{minipage}\hfill
\end{figure*}
\subsection{\textbf{SHapley Additive exPlanations (SHAP)}}
\subsubsection{Mapping SHAP Indices to Acoustic Properties}
It is critical to interpret the generic feature indices (e.g., ``Feature 1191" in \textbf{Fig. \ref{fig:shap_analysis}}) in the context of physical signal properties. 
Recall that the input to the Spectral Stability Stream is a serialized vector $\mathbf{v} \in \mathbb{R}^{L}$ constructed by horizontally stacking acoustic descriptors as shown in Equation~\ref{eq28}:
\begin{equation}
    \mathbf{v} = [\underbrace{z_1, \dots, z_T}_{\text{ZCR}}, \underbrace{e_1, \dots, e_T}_{\text{RMSE}}, \underbrace{m_{1,1}, \dots, m_{T,K}}_{\text{MFCCs}}]
    \label{eq28}
\end{equation}
where $T$ is the number of time frames and $K=40$ is the number of MFCC coefficients.

Therefore, a high SHAP value at a specific index $i$ corresponds to a specific time-step $t$ and feature type $\phi$. For an index $i$, the mapping is defined as:
\begin{equation}
    \text{Feature}(i) = 
    \begin{cases} 
      \text{ZCR at frame } i & \text{if } 0 \le i < T \\
      \text{RMSE at frame } (i - T) & \text{if } T \le i < 2T \\
      \text{MFCC}_{k} \text{ at frame } t & \text{if } i \ge 2T
   \end{cases}
\end{equation}
Consequently, the dominance of \textbf{Feature 1191} (seen in Fig. \ref{fig:shap_analysis}a) indicates that the model is heavily weighting a specific MFCC spectral coefficient around the mid-point of the utterance to identify `Fear', rather than relying solely on loudness (RMSE) or pitch roughness (ZCR).

\noindent \textbf{Clarification on Temporal Notation:} In the subsequent visualizations (Fig. \ref{fig:shap_analysis}c-d), the labels denoted as \texttt{Time\_N} refer directly to the feature indices $i$ defined above. Thus, an impact at \texttt{Time\_1191} is mathematically equivalent to the contribution of the mapped feature $\phi(1191)$ at that specific index.

\subsubsection{Global and Local Interpretability Analysis}
We employed a multi-view analysis to validate the model's decision-making logic, as presented in \textbf{Fig. \ref{fig:shap_analysis}}.

\paragraph{Global Feature Ranking (Fig. \ref{fig:shap_analysis}a)}
The mean absolute SHAP values reveal the hierarchy of acoustic cues. The dominance of features like \textbf{Feature 1191} (a high-frequency MFCC coefficient) confirms that the model prioritizes spectral texture over simple energy metrics. The distinct color coding shows that `Fear' (blue bars) is disproportionately driven by a specific subset of high-frequency spectral bands, aligning with the psychoacoustic definition of fear as a ``rough" spectral event.

\paragraph{Directional Impact (Fig. \ref{fig:shap_analysis}b)}
The Beeswarm plot resolves the \textit{directionality} of these contributions. For top features, we observe a clear separation: high feature values (red dots) result in positive SHAP scores (pushing the prediction toward `Fear'), while low values (blue dots) suppress it. This confirms the model has learned a monotonic relationship: ``more spectral roughness = higher probability of Fear."

\paragraph{Sparsity and Distribution (Fig. \ref{fig:shap_analysis}c)}
The Violin density plot highlights the \textit{sparsity} of emotional cues. Features like \textbf{Time\_1865} exhibit a ``long-tail" distribution (purple extension to the right). This indicates that this specific acoustic event is rare—most samples have near-zero impact—but when it \textit{does} occur, it is decisively predictive. This mirrors the nature of paralinguistic features (e.g., a sudden scream), which are transient rather than continuous.

\paragraph{Local Decision Decomposition (Fig. \ref{fig:shap_analysis}d)}
Finally, the Waterfall plot validates the model's \textit{additivity}. It decomposes a single prediction instance, showing how the final probability is constructed from the baseline ($E[f(x)]$) via a summation of positive (red) and negative (blue) contributions. The granular breakdown shows that the decision is not driven by a single artifact, but by the ensemble effect of hundreds of micro-features (e.g., \textbf{Time\_2367}, \textbf{Time\_1408}) distributed across the utterance.

\begin{figure*}[ht]
    \centering
    \begin{minipage}[t]{0.99\textwidth}
        \centering
        \includegraphics[width=18cm, height=9.5cm]{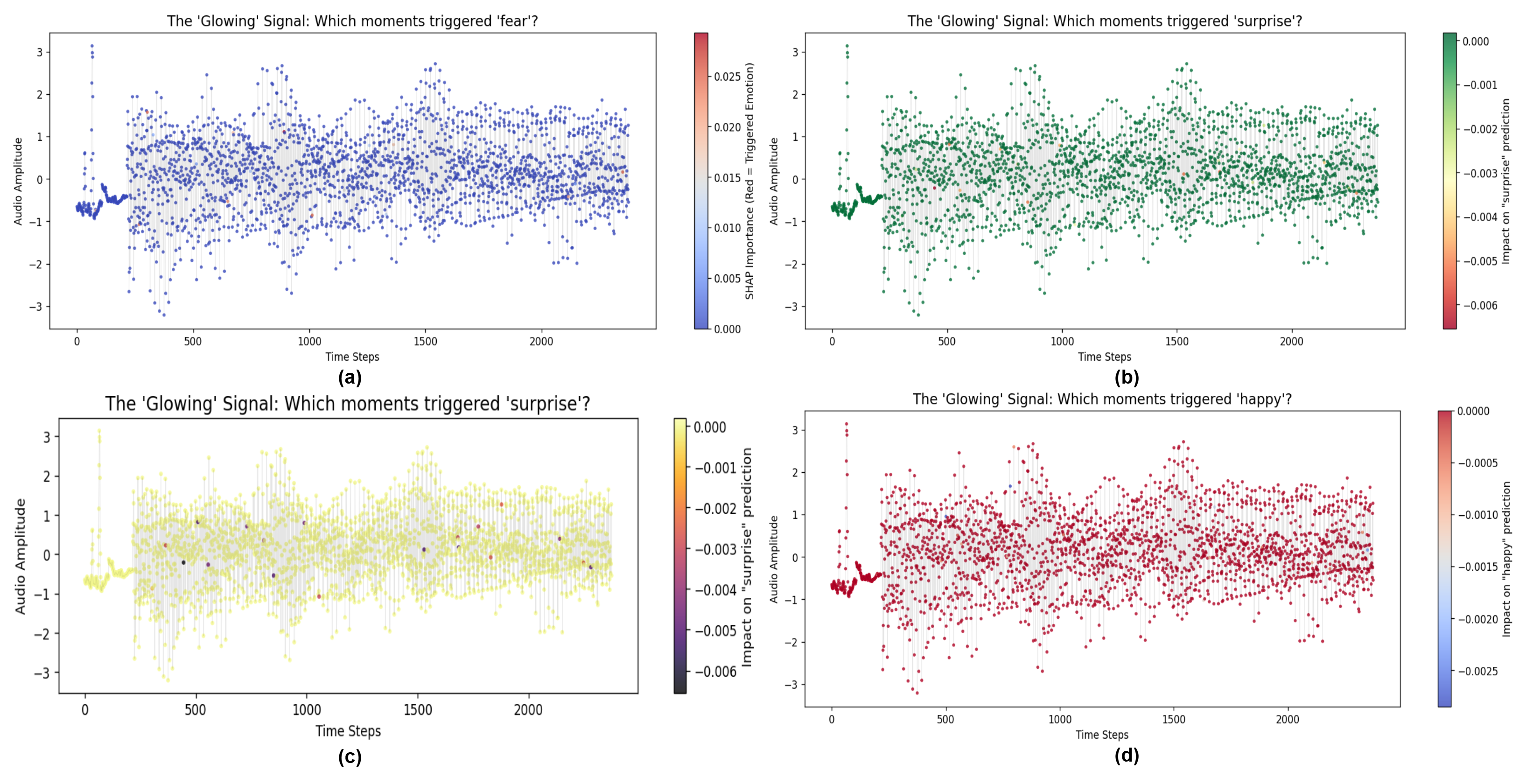}
        \caption{\textbf{Temporal SHAP Attribution (``The Glowing Signal").} These scatter plots project SHAP importance values directly onto the audio waveform for \textbf{(a)} Fear, \textbf{(b-c)} Surprise, and \textbf{(d)} Happy. Each dot represents a time-step; the color intensity indicates the magnitude of the contribution toward the target emotion prediction (Warmer/Darker colors = Strong Positive Impact).}
        \label{fig:shap_temporal}
    \end{minipage}\hfill
\end{figure*}
\vspace{1mm}
\subsubsection{Temporal Signal Attribution Analysis}
While the previous feature rankings identify \textit{what} statistics matter, \textbf{Fig. \ref{fig:shap_temporal}} visualizes \textit{when} these features occur in the physical signal. We term this the ``Glowing Signal" analysis, as it highlights the precise moments that trigger an emotional classification.

\textbf{Visual Analysis of Emotional Triggers:}
\begin{itemize}
    \item \textbf{Localized vs. Distributed Attention:} For the `Fear' class \textbf{(Fig. \ref{fig:shap_temporal}a)}, we observe high-intensity red clusters located at specific high-frequency transients. This suggests the model is identifying ``screech-like" qualities in discrete segments of the audio. In contrast, the `Happy' class \textbf{(Fig. \ref{fig:shap_temporal}d)} shows a more distributed pattern of red points across the waveform, implying that happiness is encoded in the global rhythm and continuous prosody rather than isolated shocks.
    
    \item \textbf{Onset Detection in `Surprise':} In panels \textbf{(b)} and \textbf{(c)}, the `Surprise' class demonstrates distinct activation patterns near signal onsets. The dense clustering of high-impact points (dark green/yellow) at the beginning of voiced segments aligns with the phonetic expectation that surprise is characterized by sharp, sudden vocal attacks.
    
    \item \textbf{Silence Suppression:} Across all four panels, the low-amplitude sections (silence or background noise) are predominantly covered by neutral or low-impact dots. This corroborates our Score-CAM findings, confirming that the SHAP analysis independently verifies the model's ability to ignore non-informative background noise.
\end{itemize}

\section{\textbf{Discussion}}
\label{sec:discussion}
The experimental results presented in this study highlight a fundamental limitation in current SER paradigms: while self-supervised transformers like Wav2Vec 2.0 set the benchmark for clean speech, they lack the intrinsic stability required for real-world acoustic environments. Our findings demonstrate that a hybrid fusion architecture, specifically designed to anchor deep contextual embeddings with shallow spectral features, effectively bridges this robustness gap.

\subsection{\textbf{Performance Under Acoustic Volatility}}
The detailed breakdown in \textbf{Table \ref{tab:final_results}} reveals a clear hierarchy of difficulty. Stationary noises (White, Pink, Brown) had a minimal impact on the model's performance, with accuracy remaining above 90\% even at high intensities ($\lambda=0.75$). This suggests that the model effectively learns to subtract constant background energy. However, the true test of robustness lies in non-stationary interferences like `Fish Market' and `Children Playing,' where the noise profile changes dynamically. In these scenarios, standard baselines often collapse. Our proposed model, however, maintained a commendable F1-score of 0.8921 even in the chaotic `Fish Market' environment ($\lambda=0.50$). This resilience is attributed to the \textit{Spectral Stability Stream} (Branch II), which provides a continuous stream of low-level acoustic descriptors (ZCR, MFCCs) that remain discriminative even when the transformer's phonetic attention is disrupted by overlapping speech.

\subsection{\textbf{Architectural Synergy and Trade-offs}}
The ablation study (\textbf{Table \ref{tab:ablation_study}}) provides empirical evidence for the ``complementarity hypothesis." The significant performance gap ($\Delta \approx 12\%$) between the standalone Wav2Vec model (Row B) and our Fusion Network (Row D) under noise validates the necessity of the auxiliary CNN branch. While Wav2Vec captures the \textit{linguistic} content of the emotion, the CNN branch captures the \textit{paralinguistic} texture. Crucially, the simple concatenation of these features (Row C) was insufficient to reach state-of-the-art performance. It was the introduction of \textbf{Attentive Temporal Pooling} that allowed the model to dynamically weight the importance of each stream frame-by-frame, effectively ``ignoring" corrupted segments in the temporal branch while relying on the spectral branch, and vice versa. Although this dual-stream approach incurs a slight computational cost—reducing inference speed from $\sim$80 to 75 samples/second—this 6\% reduction in throughput is a negligible trade-off for the substantial 12.35\% gain in operational reliability.

\subsection{\textbf{Interpretability as a Validation Metric}}
Finally, the quantitative success is corroborated by our Explainable AI analysis. The Score-CAM visualizations demonstrated that our model does not rely on background artifacts; instead, the attention maps tightly contour the voiced segments of the waveform. Furthermore, the Temporal SHAP analysis (Fig. \ref{fig:shap_temporal}) revealed that the model correctly identifies prosodic onsets and high-frequency transients as key emotional triggers. This alignment between the model's internal focus and human phonetic expectation confirms that the high accuracy scores are driven by genuine acoustic learning rather than statistical overfitting to the dataset.

\section{\textbf{Conclusion}}
\label{sec:conclusion}
In this work, we addressed the fragility of self-supervised SER models in non-stationary acoustic environments. While Wav2Vec 2.0 excels on clean speech, our ablation studies revealed a severe susceptibility to interference, with accuracy dropping by over 12\% under realistic noise (e.g., Railway Engine noise). We proposed a novel Dual-Stream Fusion Network combining deep contextual representations with a parallel Spectral Stability Stream (1D-CNN). This architecture achieved a test accuracy of 97.28\% on the noisy evaluation set, effectively recovering lost performance. Although the fusion mechanism introduces a marginal computational overhead (reducing inference speed to $\approx$75 samples/sec), it provides a necessary trade-off for operational robustness. Moving beyond black-box evaluation, our Explainable AI (XAI) analysis validated the model's decision-making. Score-CAM visualizations confirmed that the model aligns attention with linguistically significant voiced segments, suppressing background noise, while Temporal SHAP analysis verified that predictions rely on specific high-frequency spectral textures and prosodic onsets. These results establish the Fusion Network as a reliable candidate for real-world applications in healthcare and forensics where acoustic transparency is paramount.

\vspace{15mm}
\bibliographystyle{IEEEtran}
\bibliography{references}

@article{al2023speech,
  title={Speech emotion recognition: a comprehensive survey},
  author={Al-Dujaili, Mohammed Jawad and Ebrahimi-Moghadam, Abbas},
  journal={Wireless Personal Communications},
  volume={129},
  number={4},
  pages={2525--2561},
  year={2023},
  publisher={Springer},
  doi ={10.1007/s11277-023-10244-3}
}

@article{jordan2025speech,
  title={Speech emotion recognition in mental health: systematic review of voice-based applications},
  author={Jordan, Eric and Terrisse, Rapha{\"e}l and Lucarini, Valeria and Alrahabi, Motasem and Krebs, Marie-Odile and Descl{\'e}s, Julien and Lemey, Christophe},
  journal={JMIR mental health},
  volume={12},
  number={1},
  pages={e74260},
  year={2025},
  publisher={JMIR Publications Inc., Toronto, Canada},
  url="https://doi.org/10.2196/74260"
}

@ARTICLE{9352018,
  author={Chatterjee, Rajdeep and Mazumdar, Saptarshi and Sherratt, R. Simon and Halder, Rohit and Maitra, Tanmoy and Giri, Debasis},
  journal={IEEE Transactions on Consumer Electronics}, 
  title={Real-Time Speech Emotion Analysis for Smart Home Assistants}, 
  year={2021},
  volume={67},
  number={1},
  pages={68-76},
  doi={10.1109/TCE.2021.3056421}}

@INPROCEEDINGS{10887569,
  author={Tzeng, Jing-Tong and Leem, Seong-Gyun and Salman, Ali N. and Lee, Chi-Chun and Busso, Carlos},
  booktitle={ICASSP 2025 - 2025 IEEE International Conference on Acoustics, Speech and Signal Processing (ICASSP)}, 
  title={Noise-Robust Speech Emotion Recognition Using Shared Self-Supervised Representations with Integrated Speech Enhancement}, 
  year={2025},
  volume={},
  number={},
  pages={1-5},
  doi={10.1109/ICASSP49660.2025.10887569}}

@article{arrieta2020explainable,
  title={Explainable Artificial Intelligence (XAI): Concepts, taxonomies, opportunities and challenges toward responsible AI},
  author={Arrieta, Alejandro Barredo and D{\'\i}az-Rodr{\'\i}guez, Natalia and Del Ser, Javier and Bennetot, Adrien and Tabik, Siham and Barbado, Alberto and Garc{\'\i}a, Salvador and Gil-L{\'o}pez, Sergio and Molina, Daniel and Benjamins, Richard and others},
  journal={Information fusion},
  volume={58},
  pages={82--115},
  year={2020},
  publisher={Elsevier}
}

@article{adadi2018peeking,
  title={Peeking inside the black-box: a survey on explainable artificial intelligence (XAI)},
  author={Adadi, Amina and Berrada, Mohammed},
  journal={IEEE access},
  volume={6},
  pages={52138--52160},
  year={2018},
  publisher={IEEE}
}

@article{rudin2019stop,
  title={Stop explaining black box machine learning models for high stakes decisions and use interpretable models instead},
  author={Rudin, Cynthia},
  journal={Nature machine intelligence},
  volume={1},
  number={5},
  pages={206--215},
  year={2019},
  publisher={Nature Publishing Group UK London}
}

@ARTICLE{11205342,
  author={Chakrabarty, Sudip and Bishwas, Pappu and Bandyopadhyay, Mainak and Sublime, Jérémie},
  journal={IEEE Access}, 
  title={Can We Trust AI With Our Ears? A Cross-Domain Comparative Analysis of Explainability in Audio Intelligence}, 
  year={2025},
  volume={13},
  number={},
  pages={179733-179758},
  doi={10.1109/ACCESS.2025.3622161}}

@article{norval2025explainable,
  title={Explainable Artificial Intelligence Techniques for Speech Emotion Recognition: A Focus on XAI Models},
  author={Norval, Michael and Wang, Zenghui},
  journal={Inteligencia artificial: Revista Iberoamericana de Inteligencia Artificial},
  volume={28},
  number={76},
  pages={85--123},
  year={2025},
  publisher={Asociaci{\'o}n Espa{\~n}ola para la Inteligencia Artificial (AEPIA)}
}

@article{AOUANI2020251,
title = {Speech Emotion Recognition with deep learning},
journal = {Procedia Computer Science},
volume = {176},
pages = {251-260},
year = {2020},
note = {Knowledge-Based and Intelligent Information \& Engineering Systems: Proceedings of the 24th International Conference KES2020},
issn = {1877-0509},
doi = {https://doi.org/10.1016/j.procs.2020.08.027},
url = {https://www.sciencedirect.com/science/article/pii/S1877050920318512},
author = {Hadhami Aouani and Yassine Ben Ayed}
}

@article{baevski2020wav2vec,
  title={wav2vec 2.0: A framework for self-supervised learning of speech representations},
  author={Baevski, Alexei and Zhou, Yuhao and Mohamed, Abdelrahman and Auli, Michael},
  journal={Advances in neural information processing systems},
  volume={33},
  pages={12449--12460},
  year={2020}
}

@ARTICLE{10711189,
  author={Mohmad Dar, G. H. and Delhibabu, Radhakrishnan},
  journal={IEEE Access}, 
  title={Speech Databases, Speech Features, and Classifiers in Speech Emotion Recognition: A Review}, 
  year={2024},
  volume={12},
  number={},
  pages={151122-151152},
  doi={10.1109/ACCESS.2024.3476960}}

@article{revdess,
    doi = {10.1371/journal.pone.0196391},
    author = {Livingstone, Steven R. AND Russo, Frank A.},
    journal = {PLOS ONE},
    publisher = {Public Library of Science},
    title = {The Ryerson Audio-Visual Database of Emotional Speech and Song (RAVDESS): A dynamic, multimodal set of facial and vocal expressions in North American English},
    year = {2018},
    month = {05},
    volume = {13},
    url = {https://doi.org/10.1371/journal.pone.0196391},
    pages = {1-35},
    number = {5},

}

@article{busso2008iemocap,
  title={IEMOCAP: Interactive emotional dyadic motion capture database},
  author={Busso, Carlos and Bulut, Murtaza and Lee, Chi-Chun and Kazemzadeh, Abe and Mower, Emily and Kim, Samuel and Chang, Jeannette N and Lee, Sungbok and Narayanan, Shrikanth S},
  journal={Language resources and evaluation},
  volume={42},
  number={4},
  pages={335--359},
  year={2008},
  publisher={Springer}
}

@article{10.1145/3129340,
author = {Schuller, Bj\"{o}rn W.},
title = {Speech emotion recognition: two decades in a nutshell, benchmarks, and ongoing trends},
year = {2018},
issue_date = {May 2018},
publisher = {Association for Computing Machinery},
address = {New York, NY, USA},
volume = {61},
number = {5},
issn = {0001-0782},
url = {https://doi.org/10.1145/3129340},
doi = {10.1145/3129340},
journal = {Commun. ACM},
month = apr,
pages = {90–99},
numpages = {10}
}

@article{kaur2023trends,
  title={Trends in speech emotion recognition: a comprehensive survey},
  author={Kaur, Kamaldeep and Singh, Parminder},
  journal={Multimedia Tools and Applications},
  volume={82},
  number={19},
  pages={29307--29351},
  year={2023},
  publisher={Springer}
}

@INPROCEEDINGS{8757528,
  author={S{\"o}nmez, Ye{\c{s}}im {\"U}lgen and Varol, Asaf},
  booktitle={2019 7th International Symposium on Digital Forensics and Security (ISDFS)}, 
  title={New Trends in Speech Emotion Recognition}, 
  year={2019},
  volume={},
  number={},
  pages={1-7},
  doi={10.1109/ISDFS.2019.8757528}
}

@article{nfissi2024unveiling,
  title={Unveiling hidden factors: explainable AI for feature boosting in speech emotion recognition},
  author={Nfissi, Alaa and Bouachir, Wassim and Bouguila, Nizar and Mishara, Brian},
  journal={arXiv preprint arXiv:2406.01624},
  year={2024}
}

@inproceedings{sharma2025speech,
  title={Speech Emotion Recognition with Explainable AI: Enhancing Transparency in Emotion Recognition Systems},
  author={Sharma, Sejal and Dhingra, Aksh and Bhanot, Nirbhay and Kharb, Seema},
  booktitle={2025 4th OPJU International Technology Conference (OTCON) on Smart Computing for Innovation and Advancement in Industry 5.0},
  pages={1--6},
  year={2025},
  organization={IEEE}
}

@inproceedings{ribeiro2016should,
  title={" Why should i trust you?" Explaining the predictions of any classifier},
  author={Ribeiro, Marco Tulio and Singh, Sameer and Guestrin, Carlos},
  booktitle={Proceedings of the 22nd ACM SIGKDD international conference on knowledge discovery and data mining},
  pages={1135--1144},
  year={2016}
}

@Article{app14041553,
AUTHOR = {Kim, Tae-Wan and Kwak, Keun-Chang},
TITLE = {Speech Emotion Recognition Using Deep Learning Transfer Models and Explainable Techniques},
JOURNAL = {Applied Sciences},
VOLUME = {14},
YEAR = {2024},
NUMBER = {4},
ARTICLE-NUMBER = {1553},
URL = {https://www.mdpi.com/2076-3417/14/4/1553},
ISSN = {2076-3417},
DOI = {10.3390/app14041553}
}

@inproceedings{selvaraju2017grad,
  title={Grad-cam: Visual explanations from deep networks via gradient-based localization},
  author={Selvaraju, Ramprasaath R and Cogswell, Michael and Das, Abhishek and Vedantam, Ramakrishna and Parikh, Devi and Batra, Dhruv},
  booktitle={Proceedings of the IEEE international conference on computer vision},
  pages={618--626},
  year={2017}
}

@INPROCEEDINGS{11173395,
  author={Chakrabarty, Sudip and Bandyopadhyay, Mainak and Bishwas, Pappu},
  booktitle={2025 IEEE Guwahati Subsection Conference (GCON)}, 
  title={Memory Efficient Audio Classification Using Binarized Neural Network}, 
  year={2025},
  volume={},
  number={},
  pages={1-6},
  doi={10.1109/GCON65540.2025.11173395}}

@article{win2020,
  title={Emotion recognition system of noisy speech in real world environment},
  author={Win, Htwe Pa Pa and Khine, Phyo Thu Thu and others},
  journal={International Journal of Image, Graphics and Signal Processing (IJIGSP)},
  volume={12},
  number={2},
  pages={1--8},
  year={2020},
  doi={10.5815/ijigsp.2020.02.01}
}

@inproceedings{sc06,
  title     = {Emotion recognition in the noise applying large acoustic feature sets},
  author    = {Björn Schuller and Dejan Arsic and Frank Wallhoff and Gerhard Rigoll},
  year      = {2006},
  booktitle = {Speech Prosody 2006},
  pages     = {paper 128},
  doi       = {10.21437/SpeechProsody.2006-150},
  issn      = {2333-2042},
}

@article{article1,
author = {Xu, Mingke and Zhang, Fan and Khan, Samee and Zhang, Wei},
year = {2021},
month = {03},
pages = {1-1},
title = {Head Fusion: Improving the Accuracy and Robustness of Speech Emotion Recognition on the IEMOCAP and RAVDESS Dataset},
volume = {PP},
journal = {IEEE Access},
doi = {10.1109/ACCESS.2021.3067460}
}

@inproceedings{chak19,
  title     = {Front-End Feature Compensation and Denoising for Noise Robust Speech Emotion Recognition},
  author    = {Rupayan Chakraborty and Ashish Panda and Meghna Pandharipande and Sonal Joshi and Sunil Kumar Kopparapu},
  year      = {2019},
  booktitle = {Interspeech 2019},
  pages     = {3257--3261},
  doi       = {10.21437/Interspeech.2019-2243},
  issn      = {2958-1796},
}

@ARTICLE{10347397,
  author={Leem, Seong-Gyun and Fulford, Daniel and Onnela, Jukka-Pekka and Gard, David and Busso, Carlos},
  journal={IEEE/ACM Transactions on Audio, Speech, and Language Processing}, 
  title={Selective Acoustic Feature Enhancement for Speech Emotion Recognition With Noisy Speech}, 
  year={2024},
  volume={32},
  number={},
  pages={917-929},
  doi={10.1109/TASLP.2023.3340603}}

@ARTICLE{10018195,
  author={Yunxiang, Liu and Kexin, Zhang},
  journal={IEEE Access}, 
  title={Design of Efficient Speech Emotion Recognition Based on Multi Task Learning}, 
  year={2023},
  volume={11},
  number={},
  pages={5528-5537},
  doi={10.1109/ACCESS.2023.3237268}}

@ARTICLE{8852707,
  author={Huang, Yongming and Xiao, Jing and Tian, Kexin and Wu, Ao and Zhang, Guobao},
  journal={IEEE Access}, 
  title={Research on Robustness of Emotion Recognition Under Environmental Noise Conditions}, 
  year={2019},
  volume={7},
  number={},
  pages={142009-142021},
  doi={10.1109/ACCESS.2019.2944386}}

@article{WI0165,
title = {Robustness to noise for speech emotion classification using CNNs and attention mechanisms},
journal = {Smart Health},
volume = {19},
pages = {100165},
year = {2021},
issn = {2352-6483},
doi = {https://doi.org/10.1016/j.smhl.2020.100165},
url = {https://www.sciencedirect.com/science/article/pii/S235264832030057X},
author = {Lahiru Wijayasingha and John A. Stankovic}
}

@ARTICLE{103389,
AUTHOR={Kshirsagar, Shruti  and Pendyala, Anurag  and Falk, Tiago H. },     
TITLE={Task-specific speech enhancement and data augmentation for improved multimodal emotion recognition under noisy conditions},   
JOURNAL={Frontiers in Computer Science},   
VOLUME={Volume 5 - 2023},
YEAR={2023},
URL={https://www.frontiersin.org/journals/computer-science/articles/10.3389/fcomp.2023.1039261},
DOI={10.3389/fcomp.2023.1039261},
ISSN={2624-9898}
}

@article{chen20,
  title={Noise robust speech emotion recognition with signal-to-noise ratio adapting speech enhancement},
  author={Chen, Yu-Wen and Hirschberg, Julia and Tsao, Yu},
  journal={arXiv preprint arXiv:2309.01164},
  year={2023}
}

@inproceedings{inpr1,
author = {Zhou, Hengshun and Du, Jun and Yanhui, tu and Lee, Chin-Hui},
year = {2020},
month = {10},
pages = {4098-4102},
title = {Using Speech Enhancement Preprocessing for Speech Emotion Recognition in Realistic Noisy Conditions},
doi = {10.21437/Interspeech.2020-2472}
}

@INPROCEEDINGS{9597437,
  author={Wilf, Alex and Provost, Emily Mower},
  booktitle={2021 9th International Conference on Affective Computing and Intelligent Interaction (ACII)}, 
  title={Towards Noise Robust Speech Emotion Recognition Using Dynamic Layer Customization}, 
  year={2021},
  volume={},
  number={},
  pages={1-8},
  doi={10.1109/ACII52823.2021.9597437}}

@INPROCEEDINGS{7523189,
  author={Chenchah, Farah and Lachiri, Zied},
  booktitle={2016 2nd International Conference on Advanced Technologies for Signal and Image Processing (ATSIP)}, 
  title={Speech emotion recognition in noisy environment}, 
  year={2016},
  volume={},
  number={},
  pages={788-792},
  doi={10.1109/ATSIP.2016.7523189}}

@ARTICLE{6849440,
  author={Cao, Houwei and Cooper, David G. and Keutmann, Michael K. and Gur, Ruben C. and Nenkova, Ani and Verma, Ragini},
  journal={IEEE Transactions on Affective Computing}, 
  title={CREMA-D: Crowd-Sourced Emotional Multimodal Actors Dataset}, 
  year={2014},
  volume={5},
  number={4},
  pages={377-390},
  doi={10.1109/TAFFC.2014.2336244}}

@article{dupuis2011recognition,
  title={Recognition of emotional speech for younger and older talkers: Behavioural findings from the toronto emotional speech set},
  author={Dupuis, Kate and Pichora-Fuller, M Kathleen},
  journal={Canadian Acoustics},
  volume={39},
  number={3},
  pages={182--183},
  year={2011}
}

@article{jackson2014surrey,
  title={Surrey audio-visual expressed emotion (savee) database},
  author={Jackson, Philip and Haq, SJUoSG},
  journal={University of Surrey: Guildford, UK},
  year={2014}
}

@INPROCEEDINGS{10829485,
  author={Chatterjee, Rajdeep and Bishwas, Pappu and Chakrabarty, Sudip and Bandyopadhyay, Tathagata},
  booktitle={2024 4th International Conference on Computer, Communication, Control \& Information Technology (C3IT)}, 
  title={South Asian Sounds: Audio Classification}, 
  year={2024},
  volume={},
  number={},
  pages={1-6},
  doi={10.1109/C3IT60531.2024.10829485}}

@inproceedings{NIPS2017_8a20a862,
 author = {Lundberg, Scott M and Lee, Su-In},
 booktitle = {Advances in Neural Information Processing Systems},
 editor = {I. Guyon and U. Von Luxburg and S. Bengio and H. Wallach and R. Fergus and S. Vishwanathan and R. Garnett},
 pages = {},
 publisher = {Curran Associates, Inc.},
 title = {A Unified Approach to Interpreting Model Predictions},
 volume = {30},
 year = {2017}
}

@inproceedings{wang2020score,
  title={Score-CAM: Score-weighted visual explanations for convolutional neural networks},
  author={Wang, Haofan and Wang, Zifan and Du, Mengnan and Yang, Fan and Zhang, Zijian and Ding, Sirui and Mardziel, Piotr and Hu, Xia},
  booktitle={Proceedings of the IEEE/CVF conference on computer vision and pattern recognition workshops},
  pages={24--25},
  year={2020}
}

@article{cherry1953some,
  title={Some experiments on the recognition of speech, with one and with two ears},
  author={Cherry, Edward Collin},
  journal={Journal of the acoustical society of America},
  volume={25},
  pages={975--979},
  year={1953}
}

\end{document}